\documentclass[longauth]{aaEC}
\usepackage{graphicx}
\usepackage{natbib}
\usepackage{scalerel}

\usepackage[table]{xcolor}
\usepackage{siunitx}
\newcommand{\dd}{\mathrm{d}}

\defcitealias{Castro-EP24}{C23}
\defcitealias{2023A&A...674A.173R}{RC23}
\defcitealias{Tinker_2008}{T08}
\defcitealias{2016MNRAS.456.2486D}{D16}
\defcitealias{Ishiyama_2021}{I21}
\defcitealias{2008MNRAS.390L..64D}{D08}
\defcitealias{2011ApJ...740..102K}{K11}
\defcitealias{2013ApJ...766...32B}{B13}
\defcitealias{2014MNRAS.441.3359D}{D14}

\usepackage{txfonts}
\usepackage[pdfencoding=auto,psdextra]{hyperref}
\hypersetup{
    colorlinks=true,
    linkcolor=blue,
    filecolor=magenta,      
    urlcolor=blue,
    citecolor=blue
}
\makeatletter
\renewcommand*\aa@pageof{, page \thepage{} of \pageref*{LastPage}}
\makeatother
\usepackage[utf8]{inputenc}

\usepackage[switch, modulo]{lineno}

\usepackage{euclid}

\begin{document}

\title{\Euclid\/: Impact of halo mass-conversion model assumptions on galaxy cluster number-count analyses\thanks{This paper is published on
     behalf of the Euclid Consortium}}
\newcommand{\orcid}[1]{} 
\author{T.~Gayoux\orcid{0009-0008-9527-1490}\thanks{\email{Theo.Gayoux@obspm.fr}}\inst{\ref{aff1},\ref{aff2}}
\and P.-S.~Corasaniti\orcid{0000-0002-6386-7846}\inst{\ref{aff1},\ref{aff3}}
\and T.~R.~G.~Richardson\orcid{0000-0002-5002-7100}\inst{\ref{aff4}}
\and S.~T.~Kay\orcid{0000-0002-2277-9049}\inst{\ref{aff5}}
\and A.~M.~C.~Le~Brun\orcid{0000-0002-0936-4594}\inst{\ref{aff1}}
\and L.~Moscardini\orcid{0000-0002-3473-6716}\inst{\ref{aff6},\ref{aff7},\ref{aff8}}
\and L.~Pizzuti\orcid{0000-0001-5654-7580}\inst{\ref{aff9}}
\and S.~Bocquet\orcid{0000-0002-4900-805X}\inst{\ref{aff10}}
\and S.~Borgani\orcid{0000-0001-6151-6439}\inst{\ref{aff11},\ref{aff12},\ref{aff13},\ref{aff14},\ref{aff15}}
\and M.~Costanzi\orcid{0000-0001-8158-1449}\inst{\ref{aff11},\ref{aff13},\ref{aff12}}
\and C.~Giocoli\orcid{0000-0002-9590-7961}\inst{\ref{aff7},\ref{aff8}}
\and S.~Grandis\orcid{0000-0002-4577-8217}\inst{\ref{aff16}}
\and A.~Ragagnin\orcid{0000-0002-8106-2742}\inst{\ref{aff13}}
\and J.~Rhodes\orcid{0000-0002-4485-8549}\inst{\ref{aff17}}
\and I.~S\'aez-Casares\orcid{0000-0003-0013-5266}\inst{\ref{aff18},\ref{aff19}}
\and M.~Sereno\orcid{0000-0003-0302-0325}\inst{\ref{aff7},\ref{aff8}}
\and E.~Sarpa\orcid{0000-0002-1256-655X}\inst{\ref{aff20},\ref{aff15},\ref{aff14}}
\and B.~Altieri\orcid{0000-0003-3936-0284}\inst{\ref{aff21}}
\and A.~Amara\inst{\ref{aff22}}
\and S.~Andreon\orcid{0000-0002-2041-8784}\inst{\ref{aff23}}
\and N.~Auricchio\orcid{0000-0003-4444-8651}\inst{\ref{aff7}}
\and C.~Baccigalupi\orcid{0000-0002-8211-1630}\inst{\ref{aff12},\ref{aff13},\ref{aff14},\ref{aff20}}
\and M.~Baldi\orcid{0000-0003-4145-1943}\inst{\ref{aff24},\ref{aff7},\ref{aff8}}
\and S.~Bardelli\orcid{0000-0002-8900-0298}\inst{\ref{aff7}}
\and A.~Biviano\orcid{0000-0002-0857-0732}\inst{\ref{aff13},\ref{aff12}}
\and E.~Branchini\orcid{0000-0002-0808-6908}\inst{\ref{aff25},\ref{aff26},\ref{aff23}}
\and M.~Brescia\orcid{0000-0001-9506-5680}\inst{\ref{aff27},\ref{aff28}}
\and S.~Camera\orcid{0000-0003-3399-3574}\inst{\ref{aff29},\ref{aff30},\ref{aff31}}
\and G.~Ca\~nas-Herrera\orcid{0000-0003-2796-2149}\inst{\ref{aff32},\ref{aff33}}
\and V.~Capobianco\orcid{0000-0002-3309-7692}\inst{\ref{aff31}}
\and C.~Carbone\orcid{0000-0003-0125-3563}\inst{\ref{aff34}}
\and J.~Carretero\orcid{0000-0002-3130-0204}\inst{\ref{aff35},\ref{aff36}}
\and S.~Casas\orcid{0000-0002-4751-5138}\inst{\ref{aff37},\ref{aff38}}
\and M.~Castellano\orcid{0000-0001-9875-8263}\inst{\ref{aff39}}
\and G.~Castignani\orcid{0000-0001-6831-0687}\inst{\ref{aff7}}
\and S.~Cavuoti\orcid{0000-0002-3787-4196}\inst{\ref{aff28},\ref{aff40}}
\and K.~C.~Chambers\orcid{0000-0001-6965-7789}\inst{\ref{aff41}}
\and A.~Cimatti\inst{\ref{aff42}}
\and C.~Colodro-Conde\inst{\ref{aff43}}
\and G.~Congedo\orcid{0000-0003-2508-0046}\inst{\ref{aff44}}
\and L.~Conversi\orcid{0000-0002-6710-8476}\inst{\ref{aff45},\ref{aff21}}
\and Y.~Copin\orcid{0000-0002-5317-7518}\inst{\ref{aff46}}
\and F.~Courbin\orcid{0000-0003-0758-6510}\inst{\ref{aff47},\ref{aff48}}
\and H.~M.~Courtois\orcid{0000-0003-0509-1776}\inst{\ref{aff49}}
\and A.~Da~Silva\orcid{0000-0002-6385-1609}\inst{\ref{aff50},\ref{aff51}}
\and H.~Degaudenzi\orcid{0000-0002-5887-6799}\inst{\ref{aff52}}
\and G.~De~Lucia\orcid{0000-0002-6220-9104}\inst{\ref{aff13}}
\and H.~Dole\orcid{0000-0002-9767-3839}\inst{\ref{aff53}}
\and F.~Dubath\orcid{0000-0002-6533-2810}\inst{\ref{aff52}}
\and C.~A.~J.~Duncan\orcid{0009-0003-3573-0791}\inst{\ref{aff44}}
\and X.~Dupac\inst{\ref{aff21}}
\and S.~Dusini\orcid{0000-0002-1128-0664}\inst{\ref{aff54}}
\and S.~Escoffier\orcid{0000-0002-2847-7498}\inst{\ref{aff55}}
\and M.~Farina\orcid{0000-0002-3089-7846}\inst{\ref{aff56}}
\and R.~Farinelli\inst{\ref{aff7}}
\and S.~Farrens\orcid{0000-0002-9594-9387}\inst{\ref{aff57}}
\and F.~Faustini\orcid{0000-0001-6274-5145}\inst{\ref{aff39},\ref{aff58}}
\and S.~Ferriol\inst{\ref{aff46}}
\and F.~Finelli\orcid{0000-0002-6694-3269}\inst{\ref{aff7},\ref{aff59}}
\and M.~Frailis\orcid{0000-0002-7400-2135}\inst{\ref{aff13}}
\and E.~Franceschi\orcid{0000-0002-0585-6591}\inst{\ref{aff7}}
\and M.~Fumana\orcid{0000-0001-6787-5950}\inst{\ref{aff34}}
\and S.~Galeotta\orcid{0000-0002-3748-5115}\inst{\ref{aff13}}
\and K.~George\orcid{0000-0002-1734-8455}\inst{\ref{aff10}}
\and B.~Gillis\orcid{0000-0002-4478-1270}\inst{\ref{aff44}}
\and J.~Gracia-Carpio\inst{\ref{aff60}}
\and A.~Grazian\orcid{0000-0002-5688-0663}\inst{\ref{aff61}}
\and F.~Grupp\inst{\ref{aff60},\ref{aff62}}
\and S.~V.~H.~Haugan\orcid{0000-0001-9648-7260}\inst{\ref{aff63}}
\and W.~Holmes\inst{\ref{aff17}}
\and F.~Hormuth\inst{\ref{aff64}}
\and A.~Hornstrup\orcid{0000-0002-3363-0936}\inst{\ref{aff65},\ref{aff66}}
\and K.~Jahnke\orcid{0000-0003-3804-2137}\inst{\ref{aff67}}
\and M.~Jhabvala\inst{\ref{aff68}}
\and E.~Keih\"anen\orcid{0000-0003-1804-7715}\inst{\ref{aff69}}
\and S.~Kermiche\orcid{0000-0002-0302-5735}\inst{\ref{aff55}}
\and A.~Kiessling\orcid{0000-0002-2590-1273}\inst{\ref{aff17}}
\and M.~Kilbinger\orcid{0000-0001-9513-7138}\inst{\ref{aff57}}
\and B.~Kubik\orcid{0009-0006-5823-4880}\inst{\ref{aff46}}
\and M.~Kunz\orcid{0000-0002-3052-7394}\inst{\ref{aff70}}
\and H.~Kurki-Suonio\orcid{0000-0002-4618-3063}\inst{\ref{aff71},\ref{aff72}}
\and O.~Lahav\orcid{0000-0002-1134-9035}\inst{\ref{aff73}}
\and S.~Ligori\orcid{0000-0003-4172-4606}\inst{\ref{aff31}}
\and P.~B.~Lilje\orcid{0000-0003-4324-7794}\inst{\ref{aff63}}
\and V.~Lindholm\orcid{0000-0003-2317-5471}\inst{\ref{aff71},\ref{aff72}}
\and I.~Lloro\orcid{0000-0001-5966-1434}\inst{\ref{aff74}}
\and G.~Mainetti\orcid{0000-0003-2384-2377}\inst{\ref{aff75}}
\and D.~Maino\inst{\ref{aff18},\ref{aff34},\ref{aff19}}
\and E.~Maiorano\orcid{0000-0003-2593-4355}\inst{\ref{aff7}}
\and O.~Mansutti\orcid{0000-0001-5758-4658}\inst{\ref{aff13}}
\and S.~Marcin\inst{\ref{aff76}}
\and O.~Marggraf\orcid{0000-0001-7242-3852}\inst{\ref{aff77}}
\and M.~Martinelli\orcid{0000-0002-6943-7732}\inst{\ref{aff39},\ref{aff78}}
\and N.~Martinet\orcid{0000-0003-2786-7790}\inst{\ref{aff79}}
\and F.~Marulli\orcid{0000-0002-8850-0303}\inst{\ref{aff6},\ref{aff7},\ref{aff8}}
\and R.~J.~Massey\orcid{0000-0002-6085-3780}\inst{\ref{aff80}}
\and E.~Medinaceli\orcid{0000-0002-4040-7783}\inst{\ref{aff7}}
\and S.~Mei\orcid{0000-0002-2849-559X}\inst{\ref{aff81},\ref{aff82}}
\and Y.~Mellier\inst{\ref{aff83},\ref{aff3}}
\and M.~Meneghetti\orcid{0000-0003-1225-7084}\inst{\ref{aff7},\ref{aff8}}
\and E.~Merlin\orcid{0000-0001-6870-8900}\inst{\ref{aff39}}
\and G.~Meylan\inst{\ref{aff84}}
\and A.~Mora\orcid{0000-0002-1922-8529}\inst{\ref{aff85}}
\and M.~Moresco\orcid{0000-0002-7616-7136}\inst{\ref{aff6},\ref{aff7}}
\and E.~Munari\orcid{0000-0002-1751-5946}\inst{\ref{aff13},\ref{aff12}}
\and C.~Neissner\orcid{0000-0001-8524-4968}\inst{\ref{aff86},\ref{aff36}}
\and S.-M.~Niemi\orcid{0009-0005-0247-0086}\inst{\ref{aff32}}
\and C.~Padilla\orcid{0000-0001-7951-0166}\inst{\ref{aff86}}
\and S.~Paltani\orcid{0000-0002-8108-9179}\inst{\ref{aff52}}
\and F.~Pasian\orcid{0000-0002-4869-3227}\inst{\ref{aff13}}
\and K.~Pedersen\inst{\ref{aff87}}
\and V.~Pettorino\orcid{0000-0002-4203-9320}\inst{\ref{aff32}}
\and S.~Pires\orcid{0000-0002-0249-2104}\inst{\ref{aff57}}
\and G.~Polenta\orcid{0000-0003-4067-9196}\inst{\ref{aff58}}
\and M.~Poncet\inst{\ref{aff88}}
\and L.~A.~Popa\inst{\ref{aff89}}
\and L.~Pozzetti\orcid{0000-0001-7085-0412}\inst{\ref{aff7}}
\and F.~Raison\orcid{0000-0002-7819-6918}\inst{\ref{aff60}}
\and R.~Rebolo\orcid{0000-0003-3767-7085}\inst{\ref{aff43},\ref{aff90},\ref{aff91}}
\and A.~Renzi\orcid{0000-0001-9856-1970}\inst{\ref{aff92},\ref{aff54}}
\and G.~Riccio\inst{\ref{aff28}}
\and E.~Romelli\orcid{0000-0003-3069-9222}\inst{\ref{aff13}}
\and M.~Roncarelli\orcid{0000-0001-9587-7822}\inst{\ref{aff7}}
\and R.~Saglia\orcid{0000-0003-0378-7032}\inst{\ref{aff62},\ref{aff60}}
\and Z.~Sakr\orcid{0000-0002-4823-3757}\inst{\ref{aff93},\ref{aff94},\ref{aff95}}
\and D.~Sapone\orcid{0000-0001-7089-4503}\inst{\ref{aff96}}
\and B.~Sartoris\orcid{0000-0003-1337-5269}\inst{\ref{aff62},\ref{aff13}}
\and P.~Schneider\orcid{0000-0001-8561-2679}\inst{\ref{aff77}}
\and A.~Secroun\orcid{0000-0003-0505-3710}\inst{\ref{aff55}}
\and G.~Seidel\orcid{0000-0003-2907-353X}\inst{\ref{aff67}}
\and S.~Serrano\orcid{0000-0002-0211-2861}\inst{\ref{aff97},\ref{aff98},\ref{aff99}}
\and P.~Simon\inst{\ref{aff77}}
\and C.~Sirignano\orcid{0000-0002-0995-7146}\inst{\ref{aff92},\ref{aff54}}
\and G.~Sirri\orcid{0000-0003-2626-2853}\inst{\ref{aff8}}
\and L.~Stanco\orcid{0000-0002-9706-5104}\inst{\ref{aff54}}
\and J.-L.~Starck\orcid{0000-0003-2177-7794}\inst{\ref{aff57}}
\and J.~Steinwagner\orcid{0000-0001-7443-1047}\inst{\ref{aff60}}
\and P.~Tallada-Cresp\'{i}\orcid{0000-0002-1336-8328}\inst{\ref{aff35},\ref{aff36}}
\and A.~N.~Taylor\inst{\ref{aff44}}
\and I.~Tereno\orcid{0000-0002-4537-6218}\inst{\ref{aff50},\ref{aff100}}
\and N.~Tessore\orcid{0000-0002-9696-7931}\inst{\ref{aff101},\ref{aff73}}
\and S.~Toft\orcid{0000-0003-3631-7176}\inst{\ref{aff102},\ref{aff103}}
\and R.~Toledo-Moreo\orcid{0000-0002-2997-4859}\inst{\ref{aff104}}
\and F.~Torradeflot\orcid{0000-0003-1160-1517}\inst{\ref{aff36},\ref{aff35}}
\and I.~Tutusaus\orcid{0000-0002-3199-0399}\inst{\ref{aff99},\ref{aff97},\ref{aff94}}
\and L.~Valenziano\orcid{0000-0002-1170-0104}\inst{\ref{aff7},\ref{aff59}}
\and J.~Valiviita\orcid{0000-0001-6225-3693}\inst{\ref{aff71},\ref{aff72}}
\and T.~Vassallo\orcid{0000-0001-6512-6358}\inst{\ref{aff13}}
\and G.~Verdoes~Kleijn\orcid{0000-0001-5803-2580}\inst{\ref{aff105}}
\and A.~Veropalumbo\orcid{0000-0003-2387-1194}\inst{\ref{aff23},\ref{aff26},\ref{aff25}}
\and Y.~Wang\orcid{0000-0002-4749-2984}\inst{\ref{aff106}}
\and J.~Weller\orcid{0000-0002-8282-2010}\inst{\ref{aff62},\ref{aff60}}
\and G.~Zamorani\orcid{0000-0002-2318-301X}\inst{\ref{aff7}}
\and E.~Zucca\orcid{0000-0002-5845-8132}\inst{\ref{aff7}}
\and C.~Burigana\orcid{0000-0002-3005-5796}\inst{\ref{aff107},\ref{aff59}}
\and M.~Maturi\orcid{0000-0002-3517-2422}\inst{\ref{aff93},\ref{aff108}}
\and V.~Scottez\orcid{0009-0008-3864-940X}\inst{\ref{aff83},\ref{aff109}}
\and M.~Viel\orcid{0000-0002-2642-5707}\inst{\ref{aff12},\ref{aff13},\ref{aff20},\ref{aff14},\ref{aff15}}}
										   
\institute{Laboratoire d'etude de l'Univers et des phenomenes eXtremes, Observatoire de Paris, Universit\'e PSL, Sorbonne Universit\'e, CNRS, 92190 Meudon, France\label{aff1}
\and
Universit\'e Paris-Cit\'e, 5 Rue Thomas Mann, 75013, Paris, France\label{aff2}
\and
Institut d'Astrophysique de Paris, UMR 7095, CNRS, and Sorbonne Universit\'e, 98 bis boulevard Arago, 75014 Paris, France\label{aff3}
\and
Donostia International Physics Center (DIPC), Paseo Manuel de Lardizabal, 4, 20018, Donostia-San Sebasti\'an, Guipuzkoa, Spain\label{aff4}
\and
Jodrell Bank Centre for Astrophysics, Department of Physics and Astronomy, University of Manchester, Oxford Road, Manchester M13 9PL, UK\label{aff5}
\and
Dipartimento di Fisica e Astronomia "Augusto Righi" - Alma Mater Studiorum Universit\`a di Bologna, via Piero Gobetti 93/2, 40129 Bologna, Italy\label{aff6}
\and
INAF-Osservatorio di Astrofisica e Scienza dello Spazio di Bologna, Via Piero Gobetti 93/3, 40129 Bologna, Italy\label{aff7}
\and
INFN-Sezione di Bologna, Viale Berti Pichat 6/2, 40127 Bologna, Italy\label{aff8}
\and
Dipartimento di Fisica ``G. Occhialini", Universit\`a degli Studi di Milano Bicocca, Piazza della Scienza 3, 20126 Milano, Italy\label{aff9}
\and
University Observatory, LMU Faculty of Physics, Scheinerstr.~1, 81679 Munich, Germany\label{aff10}
\and
Dipartimento di Fisica - Sezione di Astronomia, Universit\`a di Trieste, Via Tiepolo 11, 34131 Trieste, Italy\label{aff11}
\and
IFPU, Institute for Fundamental Physics of the Universe, via Beirut 2, 34151 Trieste, Italy\label{aff12}
\and
INAF-Osservatorio Astronomico di Trieste, Via G. B. Tiepolo 11, 34143 Trieste, Italy\label{aff13}
\and
INFN, Sezione di Trieste, Via Valerio 2, 34127 Trieste TS, Italy\label{aff14}
\and
ICSC - Centro Nazionale di Ricerca in High Performance Computing, Big Data e Quantum Computing, Via Magnanelli 2, Bologna, Italy\label{aff15}
\and
Universit\"at Innsbruck, Institut f\"ur Astro- und Teilchenphysik, Technikerstr. 25/8, 6020 Innsbruck, Austria\label{aff16}
\and
Jet Propulsion Laboratory, California Institute of Technology, 4800 Oak Grove Drive, Pasadena, CA, 91109, USA\label{aff17}
\and
Dipartimento di Fisica "Aldo Pontremoli", Universit\`a degli Studi di Milano, Via Celoria 16, 20133 Milano, Italy\label{aff18}
\and
INFN-Sezione di Milano, Via Celoria 16, 20133 Milano, Italy\label{aff19}
\and
SISSA, International School for Advanced Studies, Via Bonomea 265, 34136 Trieste TS, Italy\label{aff20}
\and
ESAC/ESA, Camino Bajo del Castillo, s/n., Urb. Villafranca del Castillo, 28692 Villanueva de la Ca\~nada, Madrid, Spain\label{aff21}
\and
School of Mathematics and Physics, University of Surrey, Guildford, Surrey, GU2 7XH, UK\label{aff22}
\and
INAF-Osservatorio Astronomico di Brera, Via Brera 28, 20122 Milano, Italy\label{aff23}
\and
Dipartimento di Fisica e Astronomia, Universit\`a di Bologna, Via Gobetti 93/2, 40129 Bologna, Italy\label{aff24}
\and
Dipartimento di Fisica, Universit\`a di Genova, Via Dodecaneso 33, 16146, Genova, Italy\label{aff25}
\and
INFN-Sezione di Genova, Via Dodecaneso 33, 16146, Genova, Italy\label{aff26}
\and
Department of Physics "E. Pancini", University Federico II, Via Cinthia 6, 80126, Napoli, Italy\label{aff27}
\and
INAF-Osservatorio Astronomico di Capodimonte, Via Moiariello 16, 80131 Napoli, Italy\label{aff28}
\and
Dipartimento di Fisica, Universit\`a degli Studi di Torino, Via P. Giuria 1, 10125 Torino, Italy\label{aff29}
\and
INFN-Sezione di Torino, Via P. Giuria 1, 10125 Torino, Italy\label{aff30}
\and
INAF-Osservatorio Astrofisico di Torino, Via Osservatorio 20, 10025 Pino Torinese (TO), Italy\label{aff31}
\and
European Space Agency/ESTEC, Keplerlaan 1, 2201 AZ Noordwijk, The Netherlands\label{aff32}
\and
Leiden Observatory, Leiden University, Einsteinweg 55, 2333 CC Leiden, The Netherlands\label{aff33}
\and
INAF-IASF Milano, Via Alfonso Corti 12, 20133 Milano, Italy\label{aff34}
\and
Centro de Investigaciones Energ\'eticas, Medioambientales y Tecnol\'ogicas (CIEMAT), Avenida Complutense 40, 28040 Madrid, Spain\label{aff35}
\and
Port d'Informaci\'{o} Cient\'{i}fica, Campus UAB, C. Albareda s/n, 08193 Bellaterra (Barcelona), Spain\label{aff36}
\and
Institute for Theoretical Particle Physics and Cosmology (TTK), RWTH Aachen University, 52056 Aachen, Germany\label{aff37}
\and
Deutsches Zentrum f\"ur Luft- und Raumfahrt e. V. (DLR), Linder H\"ohe, 51147 K\"oln, Germany\label{aff38}
\and
INAF-Osservatorio Astronomico di Roma, Via Frascati 33, 00078 Monteporzio Catone, Italy\label{aff39}
\and
INFN section of Naples, Via Cinthia 6, 80126, Napoli, Italy\label{aff40}
\and
Institute for Astronomy, University of Hawaii, 2680 Woodlawn Drive, Honolulu, HI 96822, USA\label{aff41}
\and
Dipartimento di Fisica e Astronomia "Augusto Righi" - Alma Mater Studiorum Universit\`a di Bologna, Viale Berti Pichat 6/2, 40127 Bologna, Italy\label{aff42}
\and
Instituto de Astrof\'{\i}sica de Canarias, E-38205 La Laguna, Tenerife, Spain\label{aff43}
\and
Institute for Astronomy, University of Edinburgh, Royal Observatory, Blackford Hill, Edinburgh EH9 3HJ, UK\label{aff44}
\and
European Space Agency/ESRIN, Largo Galileo Galilei 1, 00044 Frascati, Roma, Italy\label{aff45}
\and
Universit\'e Claude Bernard Lyon 1, CNRS/IN2P3, IP2I Lyon, UMR 5822, Villeurbanne, F-69100, France\label{aff46}
\and
Institut de Ci\`{e}ncies del Cosmos (ICCUB), Universitat de Barcelona (IEEC-UB), Mart\'{i} i Franqu\`{e}s 1, 08028 Barcelona, Spain\label{aff47}
\and
Instituci\'o Catalana de Recerca i Estudis Avan\c{c}ats (ICREA), Passeig de Llu\'{\i}s Companys 23, 08010 Barcelona, Spain\label{aff48}
\and
UCB Lyon 1, CNRS/IN2P3, IUF, IP2I Lyon, 4 rue Enrico Fermi, 69622 Villeurbanne, France\label{aff49}
\and
Departamento de F\'isica, Faculdade de Ci\^encias, Universidade de Lisboa, Edif\'icio C8, Campo Grande, PT1749-016 Lisboa, Portugal\label{aff50}
\and
Instituto de Astrof\'isica e Ci\^encias do Espa\c{c}o, Faculdade de Ci\^encias, Universidade de Lisboa, Campo Grande, 1749-016 Lisboa, Portugal\label{aff51}
\and
Department of Astronomy, University of Geneva, ch. d'Ecogia 16, 1290 Versoix, Switzerland\label{aff52}
\and
Universit\'e Paris-Saclay, CNRS, Institut d'astrophysique spatiale, 91405, Orsay, France\label{aff53}
\and
INFN-Padova, Via Marzolo 8, 35131 Padova, Italy\label{aff54}
\and
Aix-Marseille Universit\'e, CNRS/IN2P3, CPPM, Marseille, France\label{aff55}
\and
INAF-Istituto di Astrofisica e Planetologia Spaziali, via del Fosso del Cavaliere, 100, 00100 Roma, Italy\label{aff56}
\and
Universit\'e Paris-Saclay, Universit\'e Paris Cit\'e, CEA, CNRS, AIM, 91191, Gif-sur-Yvette, France\label{aff57}
\and
Space Science Data Center, Italian Space Agency, via del Politecnico snc, 00133 Roma, Italy\label{aff58}
\and
INFN-Bologna, Via Irnerio 46, 40126 Bologna, Italy\label{aff59}
\and
Max Planck Institute for Extraterrestrial Physics, Giessenbachstr. 1, 85748 Garching, Germany\label{aff60}
\and
INAF-Osservatorio Astronomico di Padova, Via dell'Osservatorio 5, 35122 Padova, Italy\label{aff61}
\and
Universit\"ats-Sternwarte M\"unchen, Fakult\"at f\"ur Physik, Ludwig-Maximilians-Universit\"at M\"unchen, Scheinerstr.~1, 81679 M\"unchen, Germany\label{aff62}
\and
Institute of Theoretical Astrophysics, University of Oslo, P.O. Box 1029 Blindern, 0315 Oslo, Norway\label{aff63}
\and
Felix Hormuth Engineering, Goethestr. 17, 69181 Leimen, Germany\label{aff64}
\and
Technical University of Denmark, Elektrovej 327, 2800 Kgs. Lyngby, Denmark\label{aff65}
\and
Cosmic Dawn Center (DAWN), Denmark\label{aff66}
\and
Max-Planck-Institut f\"ur Astronomie, K\"onigstuhl 17, 69117 Heidelberg, Germany\label{aff67}
\and
NASA Goddard Space Flight Center, Greenbelt, MD 20771, USA\label{aff68}
\and
Department of Physics and Helsinki Institute of Physics, Gustaf H\"allstr\"omin katu 2, University of Helsinki, 00014 Helsinki, Finland\label{aff69}
\and
Universit\'e de Gen\`eve, D\'epartement de Physique Th\'eorique and Centre for Astroparticle Physics, 24 quai Ernest-Ansermet, CH-1211 Gen\`eve 4, Switzerland\label{aff70}
\and
Department of Physics, P.O. Box 64, University of Helsinki, 00014 Helsinki, Finland\label{aff71}
\and
Helsinki Institute of Physics, Gustaf H{\"a}llstr{\"o}min katu 2, University of Helsinki, 00014 Helsinki, Finland\label{aff72}
\and
Department of Physics and Astronomy, University College London, Gower Street, London WC1E 6BT, UK\label{aff73}
\and
SKAO, Jodrell Bank, Lower Withington, Macclesfield SK11 9FT, UK\label{aff74}
\and
Centre de Calcul de l'IN2P3/CNRS, 21 avenue Pierre de Coubertin 69627 Villeurbanne Cedex, France\label{aff75}
\and
University of Applied Sciences and Arts of Northwestern Switzerland, School of Computer Science, 5210 Windisch, Switzerland\label{aff76}
\and
Universit\"at Bonn, Argelander-Institut f\"ur Astronomie, Auf dem H\"ugel 71, 53121 Bonn, Germany\label{aff77}
\and
INFN-Sezione di Roma, Piazzale Aldo Moro, 2 - c/o Dipartimento di Fisica, Edificio G. Marconi, 00185 Roma, Italy\label{aff78}
\and
Aix-Marseille Universit\'e, CNRS, CNES, LAM, Marseille, France\label{aff79}
\and
Department of Physics, Institute for Computational Cosmology, Durham University, South Road, Durham, DH1 3LE, UK\label{aff80}
\and
Universit\'e Paris Cit\'e, CNRS, Astroparticule et Cosmologie, 75013 Paris, France\label{aff81}
\and
CNRS-UCB International Research Laboratory, Centre Pierre Bin\'etruy, IRL2007, CPB-IN2P3, Berkeley, USA\label{aff82}
\and
Institut d'Astrophysique de Paris, 98bis Boulevard Arago, 75014, Paris, France\label{aff83}
\and
Institute of Physics, Laboratory of Astrophysics, Ecole Polytechnique F\'ed\'erale de Lausanne (EPFL), Observatoire de Sauverny, 1290 Versoix, Switzerland\label{aff84}
\and
Telespazio UK S.L. for European Space Agency (ESA), Camino bajo del Castillo, s/n, Urbanizacion Villafranca del Castillo, Villanueva de la Ca\~nada, 28692 Madrid, Spain\label{aff85}
\and
Institut de F\'{i}sica d'Altes Energies (IFAE), The Barcelona Institute of Science and Technology, Campus UAB, 08193 Bellaterra (Barcelona), Spain\label{aff86}
\and
DARK, Niels Bohr Institute, University of Copenhagen, Jagtvej 155, 2200 Copenhagen, Denmark\label{aff87}
\and
Centre National d'Etudes Spatiales -- Centre spatial de Toulouse, 18 avenue Edouard Belin, 31401 Toulouse Cedex 9, France\label{aff88}
\and
Institute of Space Science, Str. Atomistilor, nr. 409 M\u{a}gurele, Ilfov, 077125, Romania\label{aff89}
\and
Consejo Superior de Investigaciones Cientificas, Calle Serrano 117, 28006 Madrid, Spain\label{aff90}
\and
Universidad de La Laguna, Dpto. Astrof\'\i sica, E-38206 La Laguna, Tenerife, Spain\label{aff91}
\and
Dipartimento di Fisica e Astronomia "G. Galilei", Universit\`a di Padova, Via Marzolo 8, 35131 Padova, Italy\label{aff92}
\and
Institut f\"ur Theoretische Physik, University of Heidelberg, Philosophenweg 16, 69120 Heidelberg, Germany\label{aff93}
\and
Institut de Recherche en Astrophysique et Plan\'etologie (IRAP), Universit\'e de Toulouse, CNRS, UPS, CNES, 14 Av. Edouard Belin, 31400 Toulouse, France\label{aff94}
\and
Universit\'e St Joseph; Faculty of Sciences, Beirut, Lebanon\label{aff95}
\and
Departamento de F\'isica, FCFM, Universidad de Chile, Blanco Encalada 2008, Santiago, Chile\label{aff96}
\and
Institut d'Estudis Espacials de Catalunya (IEEC),  Edifici RDIT, Campus UPC, 08860 Castelldefels, Barcelona, Spain\label{aff97}
\and
Satlantis, University Science Park, Sede Bld 48940, Leioa-Bilbao, Spain\label{aff98}
\and
Institute of Space Sciences (ICE, CSIC), Campus UAB, Carrer de Can Magrans, s/n, 08193 Barcelona, Spain\label{aff99}
\and
Instituto de Astrof\'isica e Ci\^encias do Espa\c{c}o, Faculdade de Ci\^encias, Universidade de Lisboa, Tapada da Ajuda, 1349-018 Lisboa, Portugal\label{aff100}
\and
Mullard Space Science Laboratory, University College London, Holmbury St Mary, Dorking, Surrey RH5 6NT, UK\label{aff101}
\and
Cosmic Dawn Center (DAWN)\label{aff102}
\and
Niels Bohr Institute, University of Copenhagen, Jagtvej 128, 2200 Copenhagen, Denmark\label{aff103}
\and
Universidad Polit\'ecnica de Cartagena, Departamento de Electr\'onica y Tecnolog\'ia de Computadoras,  Plaza del Hospital 1, 30202 Cartagena, Spain\label{aff104}
\and
Kapteyn Astronomical Institute, University of Groningen, PO Box 800, 9700 AV Groningen, The Netherlands\label{aff105}
\and
Infrared Processing and Analysis Center, California Institute of Technology, Pasadena, CA 91125, USA\label{aff106}
\and
INAF, Istituto di Radioastronomia, Via Piero Gobetti 101, 40129 Bologna, Italy\label{aff107}
\and
Zentrum f\"ur Astronomie, Universit\"at Heidelberg, Philosophenweg 12, 69120 Heidelberg, Germany\label{aff108}
\and
ICL, Junia, Universit\'e Catholique de Lille, LITL, 59000 Lille, France\label{aff109}}

\abstract{The large catalogues of galaxy clusters expected from the {\it Euclid} survey will enable cosmological analyses of cluster number counts that require accurate cosmological model predictions. One possibility is to use parametric fits calibrated against $N$-body simulations, which capture the cosmological parameter dependence of the halo mass function (HMF). Several studies have shown that this can be obtained through a calibration against haloes with spherical masses defined at the virial overdensity. In contrast, if different mass definitions are used for the HMF and the mass-observable relation, a mapping between them is required. Here, we investigate the impact of such a mapping on the cosmological parameter constraints inferred from galaxy cluster number counts for a fiducial $\Lambda$ cold dark matter cosmology. Using synthetic data from $N$-body simulations, we show that the standard approach, which relies on assuming a concentration-mass relation, can introduce significant systematic bias. In particular, depending on the halo mass definition and the concentration-mass relation assumed, this can lead to biased constraints at a level of more than 2$\sigma$. In contrast, we find that in all the cases we have considered, the mass conversion based on the halo sparsity statistics results in a systematic bias that is smaller than the statistical error.}
\keywords{Methods: numerical; Galaxies: clusters: general; Cosmology: large-scale structure of Universe; cosmological parameters;}

\titlerunning{\Euclid\/: The impact of halo mass conversion systematics on cluster number counts}
\authorrunning{T.~Gayoux et al.}

\maketitle
\nolinenumbers

\section{Introduction}\label{sc:Intro}

A major effort is currently being made to detect and observe large samples of galaxy clusters, which can provide unique insights into complex astrophysical phenomena that take place inside these massive structures and test the standard model of cosmology \citep[see e.g.][for reviews of the subject]{2005RvMP...77..207V,2012ARA&A..50..353K,Allen_2011}. In this perspective, observations from the \textit{Euclid} survey \citep{EuclidSkyOverview} are expected to detect several hundred thousand of clusters spanning a large range of masses and redshifts \citep{Adam-EP3}. The first sample of {\it Euclid} clusters has been presented in \citet{Q1-SP050}. Among the numerous cosmological tests that can be performed with such observations, measurements of the cluster number counts are the most promising observable to infer competitive cosmological parameter constraints \citep[see][for a forecast analysis of {\it Euclid} cluster number counts]{2016MNRAS.459.1764S}, which are complementary to those obtained from other standard probes such as galaxy clustering, weak gravitational lensing, and redshift-space distortions.

Cosmological model predictions of cluster number counts particularly depend on an accurate determination of the abundance of massive dark matter haloes, which is given by the halo mass function (HMF). In practice, this can only be obtained from the analysis of high-resolution large-volume cosmological simulations. In principle, these should be $N$-body and hydrodynamical simulations that are capable of following the clustering of dark matter as well as that of the baryonic gas, eventually leading to the formation of stars and galaxies. This is because the presence of baryons has been shown to affect the predictions of the abundance of galaxy clusters \citep[][]{2014MNRAS.442.2641V,2014MNRAS.441.1769C,2014MNRAS.439.2485C,2016MNRAS.456.2361B,2025MNRAS.537.2179K}. However, given their large computational cost, the HMF is primarily predicted using $N$-body simulations only, while the effect of baryons is accounted for in post-processing through calibrated models \citep[see e.g.][]{EP-Castro}. Even then, it is still impossible to compute the HMF for every set of cosmological parameters sampled in a cosmological data analysis. Although in recent years machine learning methods have been used to develop emulators capable of predicting the HMF using a reduced set of training simulations \citep[see e.g.][]{2016ApJ...820..108H,2020ApJ...901....5B,Saez-Casares:2024bzg}, the standard approach to circumvent this issue is to use a fitting function of the HMF calibrated against simulations. Building upon the seminal work by \citet{1974ApJ...187..425P}, the HMF is usually expressed in terms of a parametrised multiplicity function fitted against numerical estimates obtained from {\it N}-body halo catalogues \citep[see e.g.][]{Jenkins_2001,Sheth_2001,Tinker_2008,2016MNRAS.456.2486D}. To be consistent with observations, spherical halo masses are usually adopted. These are defined as the mass enclosed within a sphere for which the mean enclosed density is $\Delta$ times the mean, $\rho_{\rm m}$, or critical density, $\rho_{\rm cr}$, of the Universe. Hereafter, we consider overdensities expressed in units of the critical density $\rho_{\rm cr}$, unless specified otherwise.

The accuracy of the numerically calibrated HMF for a given halo mass definition depends primarily on the volume, the resolution of the simulations, the gravity solver, the initial conditions, and the halo detection algorithm. Furthermore, these analytical fitting functions must be able to correctly capture the cosmological dependence of the multiplicity function, the so-called non-universality of the HMF \citep{2010MNRAS.403.1353C,2011MNRAS.410.1911C,2013MNRAS.433.1230W,2020ApJ...903...87D,2022MNRAS.509.6077O}. This is because uncertainties in the modelling of the multiplicity function may result in errors that systematically affect the cosmological parameter inference from cluster number count measurements \citep[see e.g.][]{2021A&A...649A..47A,Salvati_2020}. The spherical collapse model predicts that haloes are associated with matter density perturbations that collapse at the virial overdensity, $\Delta_{\rm vir}$. This has suggested that calibrating HMF fitting functions to numerical estimates obtained from haloes with masses defined at $\Delta_{\rm vir}$ may alleviate the problem of the non-universality of the HMF, since the virial overdensity depends on the underlying cosmological model \citep[see e.g.][]{1998ApJ...495...80B,2011MNRAS.410.1911C}. Indeed, several studies have shown that this partially recovers the universality of the HMF \citep{2016MNRAS.456.2486D,2020ApJ...903...87D,2022MNRAS.509.6077O}. Nevertheless, even in such a case, discrepancies with respect to cosmological simulation predictions remain large compared to the precision expected from the upcoming generation of cluster surveys \citep[see e.g.][]{2020ApJ...903...87D}. 

To encompass these limitations, \citet[][hereafter \citetalias{Castro-EP24}]{Castro-EP24} have provided a fit of the multiplicity function that is capable of reproducing to a sub-percent level the HMF obtained using virial halo masses from a suite of $\Lambda$ cold dark matter ($\Lambda$CDM) simulations characterised by different sets of cosmological parameter values, including massive neutrinos. Hereafter, we refer to this parametrisation as the {\it Euclid}-HMF. 

Yet, the main challenge in using cluster abundances for cosmology is that individual cluster masses can be estimated only for a sub-sample of clusters. For instance, X-ray observations which capture the emission of the hot intra-cluster gas can be used to estimate the cluster mass at $\Delta=500$ under the hydrostatic equilibrium (HE) hypothesis \citep[see e.g.][]{2016A&A...594A..24P,2021ApJS..253....3H,2024OJAp....7E..13B,2024A&A...685A.106B}, while weak lensing shear profile measurements \citep[see e.g.][]{2015MNRAS.450.3665S,2019MNRAS.484.1598B,2020ApJ...890..148U,2022A&A...659A..88L} and kinematic methods \citep[][]{1997ApJ...481..633D,2006A&A...456...23B,2025A&A...693A...2S} can provide estimates of the cluster mass at $\Delta=200$. Consequently, cosmological analyses rely on scaling relations that link the cluster mass to an observable proxy. A further complication is represented by systematic bias affecting the estimated masses, for instance,  HE masses are known to be biased with respect to the total cluster mass \citep[see e.g.][for a review]{2019SSRv..215...25P}. As weak lensing observations directly probe the gravitating mass in clusters, measurements of the shear profile of clusters have become a key tool for calibrating scaling relations. In particular, in recent years cosmological analyses of cluster number count data have been performed in combination with cluster shear profile measurements to jointly constrain cosmological parameters as well as scaling relation parameters \citep[see e.g.][]{2022A&A...659A..88L,2023MNRAS.522.1601C,2024A&A...687A.178G,2024A&A...689A.298G,2024PhRvD.110h3509B,2024OJAp....7E..90C,2025A&A...695A.216K,2025PhRvD.111f3533B}.

However, weak lensing masses are not exempt from bias effects due to line-of-sight projection, halo triaxiality, miscentering, and baryonic feedback. To assess these sources of bias, numerical simulations have been used to calibrate statistical relations linking the weak lensing mass of clusters to the halo mass adopted in the definition of the HMF \citep[][]{2019ApJ...878...55B,2019MNRAS.483.2871D,2021MNRAS.505.3923S,2021MNRAS.507.5671G}. Such an approach has been extensively used in the literature. It follows that future cluster counts data analyses adopting the universal numerically calibrated HMF at the virial mass require a calibrated relation between the observable mass and the virial mass, a plan that is considered for the cosmological analyses of the upcoming \textit{Euclid} galaxy cluster data release. Instead, if two different mass definitions are used, then a mapping relating them is needed. This is the main focus of this work.

The use of mass conversion raise several important issues for the analysis of cluster number counts, including the uncertainties introduced by mass conversion models in the cosmological parameter inference, the identification of an optimal halo mass mapping that minimises systematic effects arising from mass conversion, and the extent to which these effects can impact the constraints from the cluster number counts, assuming the characteristics of the {\it Euclid} cluster sample.

Here, we address these issues. Following the work by \citet[][hereafter \citetalias{2023A&A...674A.173R}]{2023A&A...674A.173R}, we present a thorough study of the impact of the different mass conversion approaches on the cosmological parameter constraints inferred from cluster number counts data analyses. 

The paper is organised as follows: in Sect.~\ref{sec:mass_conv}, we review the general formalism presented in \citetalias{2023A&A...674A.173R} to perform the mapping of the HMF from one mass definition to another; in Sect.~\ref{sec:nbody}, we describe the numerical simulations used in our analysis; in Sect.~\ref{sec:simu}, we discuss the calibration and testing of the assumed HMF parametrisations; in Sects.~\ref{sec:mass_conv_uchuu} and~\ref{sec:mass_conv_flag}, we present the results of the cosmological parameter inference for the different mapping approaches using synthetic data from the numerical halo catalogues; finally, in Sect.~\ref{sec:conclu}, we discuss the conclusions.

\section{Halo mass conversion: general formalism}\label{sec:mass_conv}
In this section, we present the different mass conversion methods and review the mathematical formalism introduced in \citetalias{2023A&A...674A.173R}, to which we refer the reader for a detailed derivation and validation against $N$-body simulations. The simplest approach to convert from one halo mass definition to another is to assume an analytical form of the halo density profile and derive the relation between the mass at different overdensities. It is a well-established result that the density profile of haloes from $N$-body simulations is well described by the Navarro--Frenk--White (NFW) profile \citep{1997ApJ...490..493N}. This characterises the halo density profile in terms of two parameters: the halo mass $M_{\Delta}$ (defined at a given overdensity $\Delta$) and the concentration parameter $c_{\Delta}$. It was already shown by \citet{1997ApJ...490..493N} that the median concentration varies as a function of halo mass and redshift, a relation that has been extensively investigated by subsequent analyses of $N$-body haloes \citep[see e.g.][]{2015ApJ...799..108D,2019ApJ...871..168D,Ishiyama_2021}. Hence, by assuming the NFW-profile and a concentration-mass relation, it is possible to map the halo mass from one overdensity to another \citep[see e.g.][]{2003ApJ...584..702H}. We will refer to this approach as `parametric deterministic' (PD) mass conversion.

The use of a deterministic mass conversion is not unconventional in cluster cosmological data analyses. As an example, \citet{2024OJAp....7E..90C} use the HMF from \citet{2016MNRAS.456.2361B}, which was calibrated on halo catalogues from the \texttt{Magneticum} simulations\footnote{\url{http://www.magneticum.org}}  with halo masses defined at $200\,\rho_{\rm m}$. To enable comparisons at different mass definitions, \citet{2016MNRAS.456.2361B} provided fitting formulas based on the NFW profile to map the calibrated HMF from $200\,\rho_{\rm m}$ to $200\,\rho_{\rm cr}$ and $500\,\rho_{\rm cr}$. Similarly,  \citet{2021MNRAS.500.5056R} used halo catalogues from the \texttt{Magneticum} simulations to establish calibrated relations linking halo concentration and mass across various overdensities, as well as conversions between different halo mass definitions, later applied in the cosmological analysis by \citet{2025A&A...703A..25L}.

However, the relation between halo concentration and mass, and more generally between masses defined at different overdensities, is not deterministic, but rather stochastic. Numerical studies have shown that the concentration at a given halo mass exhibits a large scatter \citep[see e.g.][]{2001MNRAS.321..559B,2004A&A...416..853D,2007MNRAS.378...55M}. Hence, a more accurate mass conversion must account for the statistical nature of the mass-concentration relation. We will refer to this approach as `parametric stochastic' (PS) mass conversion, which hints at the fact that the concentration is a random variable characterised by a probability distribution function. Such an approach is certainly an improvement upon the more naive deterministic mass conversion model. 

Nonetheless, assuming NFW still imposes a strong assumption on the mass distribution within haloes, since there are deviations with respect to the NFW best-fit profile due to the halo dynamical state and distribution of substructures \citep{2014MNRAS.437.2328B,2020MNRAS.498.4450W,2022MNRAS.513.4951R}. Not surprisingly, \citet{2021MNRAS.500.5056R} found that deviations from the NFW profile induce an additional scatter on the mass conversion based on the concentration-mass relation with respect to that provided by the relation between halo masses at different overdensities, potentially affecting both PD and PS mass conversion \citep[see also][for additional scatter induced on weak lensing mass bias]{EP-Ragagnin}. So while these deviations in the density profile may be small, they lead to large changes in the mass profile \citep{2022MNRAS.513.4951R}.

As an alternative, and to avoid assuming the parametric shape of the density profile, \citet{2014MNRAS.437.2328B} proposed to quantify this shape in a non-parametric fashion by using the ratio of two halo masses defined at two distinct overdensities. Formally, given a halo with spherical halo masses $M_{\Delta_1}$ and $M_{\Delta_2}$ (defined at a given overdensity $\Delta_1$ and $\Delta_2$, respectively, with $\Delta_1<\Delta_2$), the sparsity ratio is defined as:
\begin{equation}
s_{\Delta_1,\Delta_2} =\frac{M_{\Delta_1}}{M_{\Delta_2}}\;,
\end{equation}
where $s_{\Delta_1,\Delta_2}>1$. This variate provides a proxy of the logarithmic slope of the halo mass profile\footnote{The logarithmic slope of the halo mass profile between radii enclosing the overdensities $\Delta_1$ and $\Delta_2$ reads as \citep{richardson:tel-04540773}
\begin{equation}
\gamma_{\Delta_1,\Delta_2}\equiv\frac{\ln{M_{\Delta_1}-\ln{M_{\Delta_2}}}}{\ln{R_{\Delta_1}}-\ln{R_{\Delta_2}}}= \frac{3\ln{s_{\Delta_1,\Delta_2}}}{\ln{\left(\frac{\Delta_1}{\Delta_2}s_{\Delta_1,\Delta_2}\right)}}\;.
\end{equation}}, alternatively it can be interpreted as a measurement of the excess mass $\Delta{M}_{12}$ enclosed in the shell comprised between the radii $r_{\Delta_1}$ and $r_{\Delta_2}$ relative to the inner halo mass $M_{\Delta_2}$, i.e. $s_{\Delta_1,\Delta_2}=\Delta{M}_{12}/M_{\Delta_2}+1$. It is worth remarking that given $\Delta_1 < \Delta_2$, the particular values of $\Delta_1$ and $\Delta_2$ are otherwise unconstrained. As such, the halo mass profile can be characterised by an arbitrarily large set of sparsity ratios and thus provides a non-parametric statistical description of the halo mass distribution.
 
Indeed, the statistics of these ratios has been shown to carry cosmological \citep{2018ApJ...862...40C,2021ApJ...911...82C,2022MNRAS.516..437C,2023A&A...674A.173R} and astrophysical \citep{2022MNRAS.513.4951R} information encoded in the cluster mass profile beyond the NFW-concentration. 
In particular, \citet{2022MNRAS.516..437C} showed that the bulk of the cosmological information encoded in the halo mass profile is contained in different sparsity configurations, namely those that are minimally correlated. In stark contrast, assuming the NFW profile compresses this information into only two parameters, the halo mass and concentration, thereby fully determining the mass at any overdensity radius. As a consequence, different NFW-based sparsity configurations are highly correlated, in tension with the weaker correlations among sparsities measured in $N$-body halo catalogues \citep[see figures 3 and 4 in][]{2022MNRAS.516..437C}. Because of this, the use of sparsities can provide a more general framework to convert halo masses.

In their recent work, \citetalias{2023A&A...674A.173R} present a mathematical formalism allowing one to convert the HMF from one mass definition to the other using sparsity statistics. We will refer to this approach as `non-parametric stochastic' (NPS) mass conversion. Moreover, \citetalias{2023A&A...674A.173R} show that since in the case of the NFW profile a single sparsity can be mapped onto the concentration parameter, the same formalism can be used to investigate the parametric stochastic mass conversion as well as the parametric deterministic one. Here, we present a study of the impact of the different mass conversion approaches on the cosmological parameter constraints inferred from cluster number counts data analyses. 

\subsection{Non-parametric stochastic (NPS)}
Hereafter, we will briefly review the mathematical formalism introduced in \citetalias{2023A&A...674A.173R}, to which we refer the reader for a detailed derivation of the formalism and validation against $N$-body simulations.

Let us consider a population of haloes with spherical overdensity masses, $M_{\Delta_1}$. These masses can be thought of as being drawn from the density function $p_{\Delta_1}(M_{\Delta_1})\equiv \dd{n}/\dd{M}_{\Delta_1}$; that is, the HMF at $M_{\Delta_1}$. Let us also consider the masses $M_{\Delta_2}$ of the same halo population, but defined at another overdensity $\Delta_2>\Delta_1$. These can be thought of as being drawn from the density function $p_{\Delta_2}(M_{\Delta_2})\equiv \dd{n}/\dd{M}_{\Delta_2}$. To map one mass function into another, we require knowledge of the stochastic relation between the two mass definitions, which is given by the sparsity $s_{\Delta_1,\Delta_2}=M_{\Delta_1}/M_{\Delta_2}$, which we assume to be drawn from the conditional probability density function, $p_{\rm s}(s_{\Delta_1,\Delta_2}|M_{\Delta_1})$.

We now have all of the ingredients needed to derive the mapping between the two HMFs using the transformation rules of random variables. More specifically, given two pairs of random variables $(X\equiv M_{\Delta_1},Y\equiv s_{\Delta_1,\Delta_2})$ and $(Z\equiv M_{\Delta_2},W\equiv s_{\Delta_1,\Delta_2})$, we want to find the mapping between their joint probability density functions, knowing that these variables are related by 
\begin{equation}
\left\{ 
\begin{aligned} 
  X &= Z\, W\\
  Y &= W
\end{aligned} \;,
\right.
\end{equation}
whose Jacobian reads as
\begin{equation}
    J\equiv  \begin{vmatrix} \partial_Z{X} & \partial_W{X} \\ \partial_Z{Y} & \partial_W{Y} \end{vmatrix} =   \begin{vmatrix} W & Z \\ 0 & 1 \end{vmatrix}= \abs{W} = s_{\Delta_1,\Delta_2}\;.
\end{equation}
From this, the joint probability density functions of the two pairs of random variables $p_1$ and $p_2$ are related by
\begin{equation}\label{eqn:joint_m2s12}
    p_{2}\left(M_{\Delta_2},s_{\Delta_1,\Delta_2}\right)=p_{1}\left(M_{\Delta_1},s_{\Delta_1,\Delta_2}\right)\, s_{\Delta_1,\Delta_2}\;.
\end{equation}
Using the fact that the HMF at $M_{\Delta_2}$ can be obtained by marginalising Eq.~(\ref{eqn:joint_m2s12}) over the distribution of sparsities, and given the fact that the joint probability distribution on the right-hand side can be decomposed as $p_{1}\left(M_{\Delta_1},s_{\Delta_1,\Delta_2}\right)=p_{\rm s}\left(s_{\Delta_1,\Delta_2}|M_{\Delta_1}\right)p_{\Delta_1}(M_{\Delta_1})$, we finally obtain the relation that maps $\dd{n}/\dd{M}_{\Delta_1}$ into $\dd{n}/\dd{M}_{\Delta_2}$

\begin{equation}\label{eq:nonparamstoc_conv}
    \frac{\dd{n}}{\dd{M}_{\Delta_2}}=\int_{1}^{\infty} s_{\Delta_1,\Delta_2}\, p_{\rm s}\left(s_{\Delta_1,\Delta_2}| M_{\Delta_1}\right)\frac{\dd{n}\left(M_{\Delta_1}\right)}{\dd{M}_{\Delta_1}}\dd{s}_{\Delta_1,\Delta_2}\;,
\end{equation}
where $M_{\Delta_1}=s_{\Delta_1,\Delta_2}M_{\Delta_2}$. We note that Eq.~(\ref{eq:nonparamstoc_conv}) has been derived without making any assumption on the underlying shape of the halo density profile. As a result, marginalising over the conditional distribution of sparsities, Eq.~(\ref{eq:nonparamstoc_conv}) allows us to correctly propagate all possible variations in the mass profile between $\Delta_1$ and $\Delta_2$ that are present in the halo sample. 

\subsection{Parametric stochastic (PS)}
Let us now consider the case in which we assume that the halo radial density profile follows exactly the NFW profile. In such a case, let us consider a sample of haloes with masses $M_{\Delta_1}$ drawn from ${\rm d}n/{\rm d}M_{\Delta_1}$ and concentration parameters $c_{\Delta_1}$ drawn from $p_{\rm c}(c_{\Delta_1}|M_{\Delta_1})$, where the latter can be obtained from simulations by computing the distribution of concentrations best-fitting the density profile of haloes with mass $M_{\Delta_1}$. Since for a given pair of values of $(M_{\Delta_1},c_{\Delta_1})$, one can compute the halo mass at any other overdensity $\Delta_2> \Delta_1$, it is possible to compute the corresponding NFW sparsity, $s^{\rm NFW}_{\Delta_1,\Delta_2}$. As shown by \citetalias{2023A&A...674A.173R}, this implies a continuous differentiable relation between concentration and sparsity, $s^{\rm NFW}_{\Delta_1,\Delta_2} = f_{\rm s}(c_{\Delta_1})$, and its inverse $c_{\Delta_1} = f_{\rm c}(s^{\rm NFW}_{\Delta_1,\Delta_2})$, such that the conditional probability density function of the NFW sparsity can be expressed in terms the conditional distribution of concentrations, $p_c(c_{\Delta_1}|M_{\Delta_1})$,
\begin{equation}\label{eqn:nfw_sparsity_concentration_stat}
    p_{\rm s}\left(s^{\rm NFW}_{\Delta_1,\Delta_2}| M_{\Delta_1}\right) = p_{\rm c}\left(f_c(s^{\rm NFW}_{\Delta_1,\Delta_2})|M_{\Delta_1}\right)\left|\frac{\dd{f}_c}{\dd{s}}\left(s^{\rm NFW}_{\Delta_1,\Delta_2}\right)\right|\;,
\end{equation}
where the conditional distribution of the concentration parameter is usually modelled as a log-normal density function with a mean specified by a given concentration-mass relation and a given width.
Using Eq.~(\ref{eqn:nfw_sparsity_concentration_stat}), which is the mapping between the statistics of the concentration parameter and the NFW-sparsity, in combination with Eq.~(\ref{eq:nonparamstoc_conv}) we can derive the halo mass conversion from $M_{\Delta_1}$ to $M_{\Delta_2}$ under the more restrictive assumption that the halo density profile is described by the NFW-profile, while accounting for the statistical distribution of the halo concentrations. 

\subsection{Parametric deterministic (PD)}
The standard approach \citep{2003ApJ...584..702H} to convert the HMF from one mass definition to another is similar to the PS approach described above, but makes the additional assumption that the scatter around the concentration mass relation can be neglected. As such, this is equivalent to assuming the conditional distribution of concentrations is given by
\begin{equation}
    p_{\rm c}\left(c_{\Delta_1}|M_{\Delta_1}\right) = \delta_{\rm D}\left[c_{\Delta_1}-\bar{c}_{\Delta_1}(M_{\Delta_1})\right]\;,
\end{equation}
where $\delta_{\rm D}(x)$ is the Dirac-delta function and $\bar{c}_{\Delta_1}(M_{\Delta_1})$ is the assumed (median) concentration-mass relation calibrated from simulations.

\section{\textit{N}-body simulations}\label{sec:nbody}
Here, we list the simulation datasets used to perform the numerical analyses.

\subsection{Uchuu halo catalogues}
We used the halo catalogues from the Uchuu $N$-body simulation suite \citep[][hereafter \citetalias{Ishiyama_2021}]{Ishiyama_2021}. These are dark matter-only simulations of a flat $\Lambda$CDM model with parameters set to the best-fit {\it Planck} 2015 cosmology \citep{Planck2016}: $\Omega_{\rm m} = 0.3089$, $\Omega_{\rm b} = 0.0486$, $h = 0.6774$, $n_{\rm s} = 0.9667$, $\sigma_8 = 0.8159$. These simulations were realised with the \texttt{GreeM}  \citep{2009PASJ...61.1319I,2012arXiv1211.4406I} code that implements a TreePM gravity solver. Specifically, we used the halo catalogues from the simulation of $(2\,h^{-1}\,{\rm Gpc})^3$ volume, the largest simulated box of the Uchuu suite, with $12\,800^3$ $N$-body particles corresponding to a tracer particle mass $m_{\rm p}=3.3\times 10^8\,h^{-1}\,M_{\odot}$ and gravitational softening length is $4.27\,h^{-1}$kpc. These catalogues were generated using the \texttt{ROCKSTAR} halo finder \citep{2013ApJ...762..109B} in $24$ redshift snapshots in the range $0\le z \le 2$. Finally, spherical overdensity masses $M_{\rm vir}$, $M_{200}$, and $M_{500}$ were computed for all detected haloes, with $M_{\rm vir}$ estimated using both bound and unbound particles. It is worth remarking that the lowest-mass halo considered in our analysis contains more than $10^5$ particles, well above the threshold of sensitivity of the halo finder to numerical effects \citep[see e.g.][for numerical tests on \texttt{ROCKSTAR}]{2011MNRAS.415.2293K,2013ApJ...762..109B,2022A&A...664A..42V,2024MNRAS.527.5603M}. We considered this particular simulation, since it provides an optimal benchmark to test for any systematic uncertainty introduced by the halo mass conversion method. First of all, the large-volume and high-mass resolution guarantees an accurate determination of the HMF, especially at the high-mass end. Secondly, it probes a volume of the same size as the PICCOLO simulations, which were used to calibrate the {\it Euclid}-HMF in \citetalias{Castro-EP24}. In fact, these are simulations covering a ($2\,h^{-1}\,{\rm Gpc}$)$^3$ volume with $4\times 1280^3$ particles generated with the OpenGadget code that also uses a TreePM-based gravity solver. Furthermore, the halo catalogues used for the {\it Euclid}-HMF calibration were also generated using the \texttt{ROCKSTAR} halo finder. Hence, the use of the Uchuu halo catalogues in our analysis allows us to test the {\it Euclid}-HMF against a simulation that was not used in the calibration. Moreover, it enables us to perform a re-calibration of the fitting function adopted for the {\it Euclid}-HMF with the intent of having an analytical fit that reproduces Uchuu's numerical estimates within Poisson errors. We refer to this recalibrated HMF as {\it Euclid}\textbackslash Uchuu-HMF, which provides us with an accurate reference for testing potential systematic effects due to the mass conversion. 
We refer the reader to Appendix~\ref{HMFcalibration} for a detailed description of the HMF parametrisation, the fitting procedure, and the best-fit parameter values. 

\subsection{Flagship light-cone halo catalogues}\label{sec:flagship}
In order to perform a forecast analysis of the impact of halo mass conversion models on the cosmological parameter inference of {\it Euclid}-like number count estimates, we used the light-cone halo catalogues from the Flagship $N$-body simulation \citep{EuclidSkyFlagship}, which are publicly available on CosmoHub\footnote{\url{https://cosmohub.pic.es/catalogs/352}} \citep[][]{2017ehep.confE.488C,TALLADA2020100391}. The simulation consisting of a $(3\,h^{-1}{\rm Gpc})^3$ volume with $16\,000^3$ particles (corresponding to $m_{\rm p}=10^9\,h^{-1}\,M_{\odot}$) for a $\Lambda$CDM model with cosmological parameters set to the {\it Euclid} fiducial values: $\Omega_{\rm m} = 0.319$, $\Omega_{\rm b} = 0.049$, $h = 0.67$, $n_{\rm s} = 0.96$, $\sigma_8 = 0.813$, $\Omega_{\rm r} = 0.00005509$, $\Omega_\nu = 0.00140343$, $\wde=-1.0$. This was run with the \texttt{PKDGRAV3} code, which implements a TreePM-based gravity solver with a Fast Multiple Method \citep{2016asclsoft09016P}. Halo catalogues were generated using the \texttt{ROCKSTAR} halo finder. Specifically, we used the WIDE light-cone dataset, corresponding to one octant of the sky within the solid angle $\Delta\Omega$ defined by the angular coordinates $0 \leq {\rm RA}/{\rm deg} \leq 90$ and $0 \leq {\rm Dec}/{\rm deg} \leq 90$ covering the interval redshift  $0\le z \le3$.

\begin{figure*}[t]
\centering
\includegraphics[width = 0.8\linewidth]{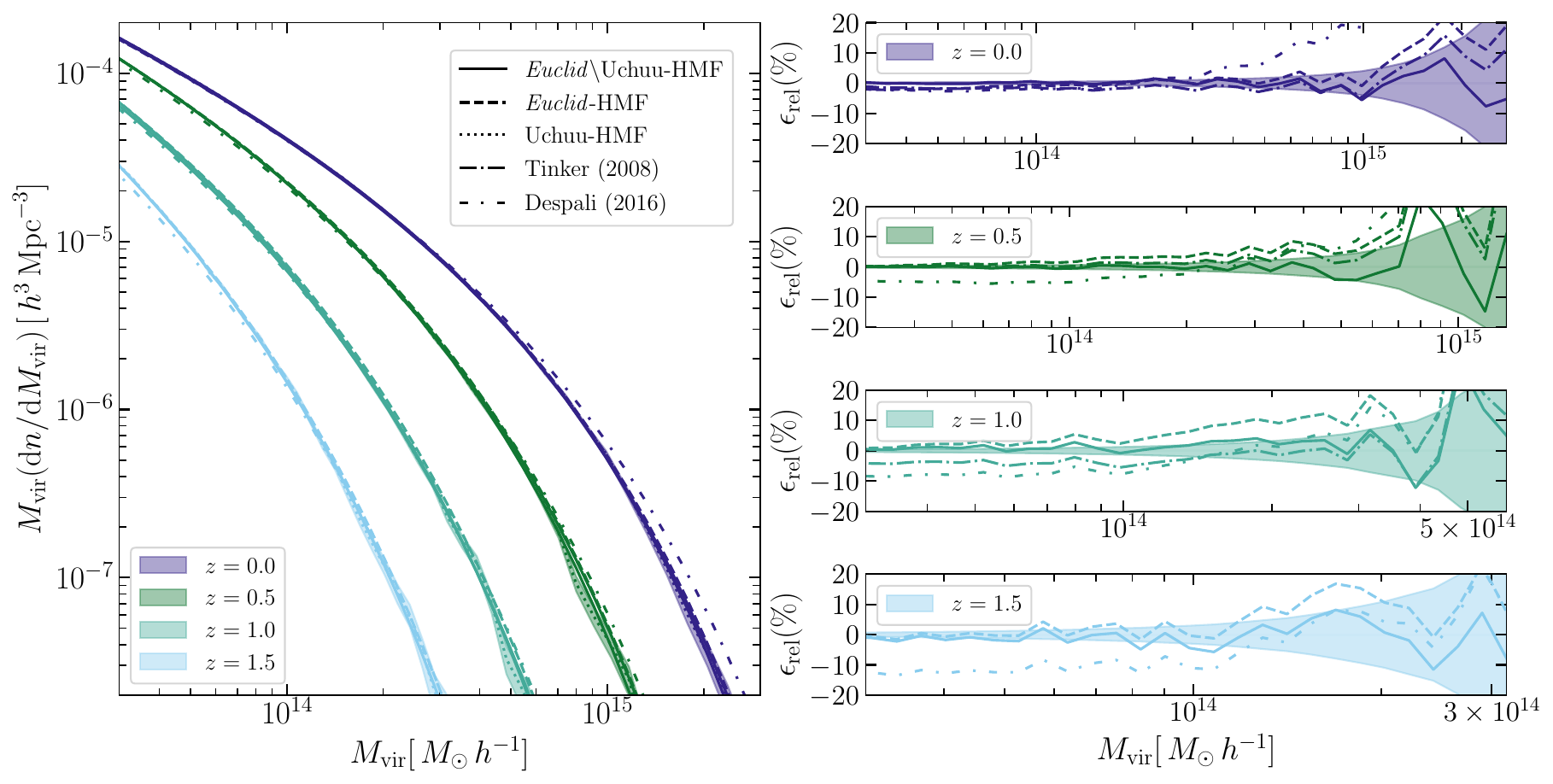}
\caption{\textit{Left panel}: HMF from the Uchuu halo catalogues with virial masses at $z=0.0$ (dotted purple line), $0.5$ (dotted green line), $1.0$ (dotted dark turquoise line), and $1.5$ (dotted light blue line) against the recalibrated {\it Euclid}-HMF (solid lines), the {\it Euclid}-HMF predictions (short-dashed lines), the HMF parametrisation from \citetalias{Tinker_2008} (dash-dotted lines), and \citetalias{2016MNRAS.456.2486D} (short-dash-dotted lines). The different colours correspond to the various redshift snapshots. \textit{Right panels}: Relative difference with respect to the Uchuu-HMF from $z=0$ (top) to $1.5$ (bottom). The shaded area corresponds to the Poisson errors.}
\label{fig:uchuu_cali}
\end{figure*}

\begin{figure*}[ht]
\centering
\includegraphics[width = 0.9\textwidth]{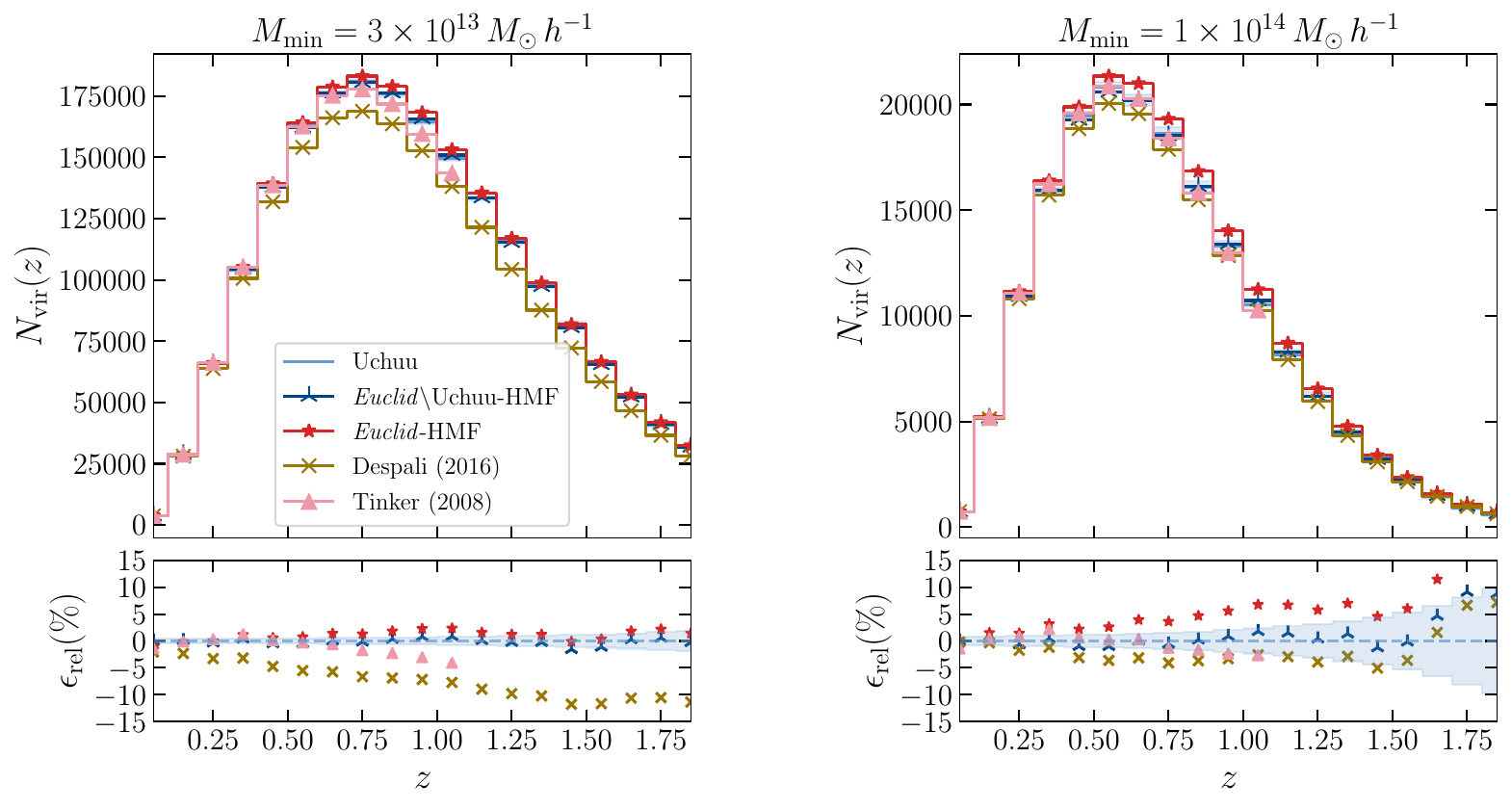}
\caption{Number counts for haloes with virial masses $M_{\rm vir}\ge 3\times 10^{13}\,M_{\odot}\,h^{-1}$ ({\it left panel}) and $M_{\rm vir}\ge 10^{14}\,M_{\odot}\,h^{-1}$ ({\it right panel}) as function of redshift from the Uchuu data (solid light blue line) against the predictions obtained assuming {\it Euclid}\textbackslash Uchuu-HMF (solid dark blue line with tri marker), {\it Euclid}-HMF (solid red line with star marker), \citetalias{2016MNRAS.456.2486D} (solid goldenrod line with cross marker) and \citetalias{Tinker_2008} (solid pink line with triangle marker). The bottom panels show the relative differences with respect to the Uchuu data. The shaded area corresponds to the Poisson errors.}
\label{fig:uchuu_ncount_vir}
\end{figure*}

\section{HMF calibration and validation}\label{sec:simu}
In this section, we describe the estimation of the numerical HMFs from the Uchuu halo catalogues, while we present in Appendix~\ref{ConditionalSparsityCalibration} the determination and calibration of the conditional sparsity density functions. These will be used to test the effect of the mass conversion models on the cluster number counts.

\subsection{Numerical HMF estimation}
We estimated the numerical HMF from the Uchuu halo catalogue at a given redshift, $z$, and for a given halo mass definition,  $M$, as
\begin{equation}
\frac{{\rm d}n}{{\rm d}\ln{M}}(M,z)= \frac{1}{V_{\rm box}}\frac{N(M,z)}{\Delta\!\ln{M}}\;,
\end{equation}
where $V_{\rm box}$ is the simulation volume and $N(M,z)$ is the number of haloes with mass $M$ in logarithmic mass bins of size $\Delta\!\ln{M}$ at redshift $z$. The size of the bins was set so as to guarantee that the highest-mass bin contains at least ten haloes, which ensures that the relative error on the HMF at the high-mass end is approximately smaller than $30\%$. To this purpose, the value of ${\Delta}\!\ln{M}$ was found iteratively for each mass definition and redshift. We estimated the Poisson error in each bin as
\begin{equation}\label{shot_noise}
    \sigma_{{\rm d}n/{\rm d}\ln{M}} =\frac{1}{V_{\rm box}}\frac{\sqrt{N(M,z)}}{\Delta\!\ln{M}}\;.
\end{equation}
We computed the HMF at $M_{\rm vir}$, $M_{200}$, and $M_{500}$ for all the $24$ redshift snapshots in the range $0\le z\le 2$. We refer to these numerical estimates as the Uchuu-HMF.

\subsection{Numerical HMF versus fitting functions}\label{sec:resolution_effects}
We compare the Uchuu-HMF data at $ z\in\{0, 0.5, 1, 1.5\}$ against the {\it Euclid}-HMF and {\it Euclid}\textbackslash Uchuu-HMF fits, as well as the predictions obtained using standard HMF parametrisations from \citet[][hereafter \citetalias{Tinker_2008}]{Tinker_2008} and \citet[][hereafter \citetalias{2016MNRAS.456.2486D}]{2016MNRAS.456.2486D}. These are shown in Fig.~\ref{fig:uchuu_cali} (left panel) together with the relative differences with respect to the Uchuu-HMF data (right panels), where the shaded areas correspond to the Poisson errors. Notice that in the case of the HMF from \citetalias{Tinker_2008}, the validity of the parametrisation is limited to $z\le 1$; hence, we do not show the comparison at $z=1.5$. 

We can see that in the case of the HMF from \citetalias{2016MNRAS.456.2486D} there are systematic differences of order of $15\%$ at $M_{\rm vir}<10^{14}\,M_{\odot}\,h^{-1}$ at $z=1.5$, while for larger masses the predictions are within the Poisson errors. At $z=0$, these discrepancies reduce to $\sim 2\%$, while at the high-mass end they increase up to $\sim 20\%$. These results are consistent with a similar comparison presented in \citetalias[][]{Ishiyama_2021}. We can see that systematic differences also occur in the case of the HMF from \citetalias{Tinker_2008}. These systematics can reach $\sim 5\%$ at $z=1$ for $M_{\rm vir}\lesssim 10^{14}\,M_{\odot}\,h^{-1}$, while at higher masses the differences remain within the numerical uncertainties. Such discrepancies result from the different resolutions and volumes of the simulations used to calibrate the \citetalias{Tinker_2008} and \citetalias{2016MNRAS.456.2486D} parametrisations. 

Let us now consider the {\it Euclid}-HMF and {\it Euclid}\textbackslash Uchuu-HMF fits. We can see that the latter reproduces the numerical results well within the Poisson errors across the whole range of masses and redshifts. This is not surprising since the fitting parameters have been calibrated to the Uchuu results. Nonetheless, it shows the ability of the parametrisation to capture the shape of the numerically estimated HMF within the statistical uncertainties. We may also notice that the {\it Euclid}-HMF reproduces quite well the Uchuu data. This is particularly evident at $z=0$ and $M_{\rm vir}\gtrsim 10^{14}\,M_{\odot}\,h^{-1}$, where deviations remain below the Poisson noise. Nonetheless, we note larger discrepancies above Poisson errors at higher redshifts. In particular, at $z=1$ deviations can be as large as $10\%$ for virial masses $M_{\rm vir}\gtrsim 10^{14}\,M_{\odot}\,h^{-1}$. Over the same mass range, these deviations decrease below the $10\%$ level at $z=0.5$ and $z=0$, but also increase above Poisson errors at the low-mass end, $M_{\rm vir}\lesssim 10^{14}\,M_{\odot}\,h^{-1}$. These discrepancies may be a manifestation of the lower mass resolution of the PICCOLO simulations ($4.4\times 10^{10}\le m_{\rm p}[h^{-1}M_{\odot}]\le 10.8\times 10^{10}$) compared to that of the Uchuu run. In any case, given that the {\it Euclid}-HMF slightly overestimates the Uchuu-HMF at the high-mass end, we may expect this to cause small differences in the expected number counts up to redshift $\sim 1$, which we investigate next.

\subsection{Number count validation}
Here, we compare the halo number counts as a function of redshift computed using the Uchuu-HMF to those obtained from the {\it Euclid}\textbackslash Uchuu-HMF and {\it Euclid}-HMF fitting functions, respectively, and we refer the readers to Appendix~\ref{ClusterCounts} for a description of the standard formula involved in the computation of the number counts. We consider the case of a survey with a sky coverage of $15\,000$ deg$^2$ and with haloes selected in bins of size $\Delta{z}=0.1$. For simplicity, we consider a selection in mass with two different mass cuts: 1) one including galaxy group-size haloes, $M_{\rm vir}\geq 3\times 10^{13}\,M_{\odot}\,h^{-1}$; 2) one limited to massive clusters, $M_{\rm vir}\geq 10^{14}\,M_{\odot}\,h^{-1}$. 

In Fig.~\ref{fig:uchuu_ncount_vir}, we plot in the upper panels, the expected number counts from the Uchuu-HMF at $M_{\rm vir}$ (solid light blue line) against the predictions of the {\it Euclid}-HMF (red line with star marker) and {\it Euclid}\textbackslash Uchuu-HMF (dark blue line with tri marker) fits as well as those obtained assuming the HMF parametrisation by \citetalias{Tinker_2008} (pink line with triangle marker) and \citetalias{2016MNRAS.456.2486D} (goldenrod line with cross marker) for the lowest (left panels) and highest (right panels) mass cuts, respectively. We show the relative difference with respect to the Uchuu's expected counts in the lower panels. As already mentioned in Sect.~\ref{sec:resolution_effects}, the counts are limited up to $z=1$ in the case of \citetalias{Tinker_2008}. We find that the {\it Euclid}-HMF slightly overestimates the number counts for both mass cuts, while the {\it Euclid}\textbackslash Uchuu-HMF predictions are in good agreement with the Uchuu counts. In the case of the {\it Euclid}-HMF, differences do not exceed a level of $3\%$ and remain constant within Poisson errors over the full redshift interval for the lowest-mass cut, while they increase above Poisson errors for the high-mass cut, though they are still limited to the $5$--$7\%$ level in the range $0.8\lesssim z\lesssim 1.3$. In contrast, we can see that the number counts predicted by the \citetalias{2016MNRAS.456.2486D} fitting function deviate by more than $5\%$ over the whole redshift interval for the lowest-mass cut. In the case of the highest-mass cut, deviations are nearly constant at a level of $3$--$5\%$ up to $z\sim 1$, though still above Poisson noise. The predicted counts from the HMF by \citetalias{Tinker_2008} are in slightly better agreement with the Uchuu data compared to the \citetalias{2016MNRAS.456.2486D} predictions. In such a case, deviations occur at $z\gtrsim 0.7$ and reach $\sim 5\%$ at $z=1$ for the lowest-mass cut, while in the case of the higher-mass cut, differences are within the Poisson errors up to $z\simeq 0.9$. Such discrepancies are a direct consequence of the differences between the HMF predictions discussed in the previous section.

\section{Testing mass conversion models}\label{sec:mass_conv_uchuu}
We used the Uchuu-HMF numerical estimates at $M_{200}$ and $M_{500}$ to compute the corresponding number counts as described in Appendix~\ref{ClusterCounts}. These counts provide a benchmark dataset to test the validity of the different mass conversion models presented in Sect.~\ref{sec:mass_conv}.
Hereafter, we will refer to the PS mass conversion performed using the numerical distributions of halo concentrations from the Uchuu catalogues as PS-num, while we will denote as PS-theo the mass conversion performed assuming a log-normal approximation of the distributions of halo concentrations. We refer the reader to Appendix~\ref{ConditionalSparsityCalibration} for a detailed discussion.

\begin{figure*}[ht]
\centering
\includegraphics[width = 0.9\linewidth]{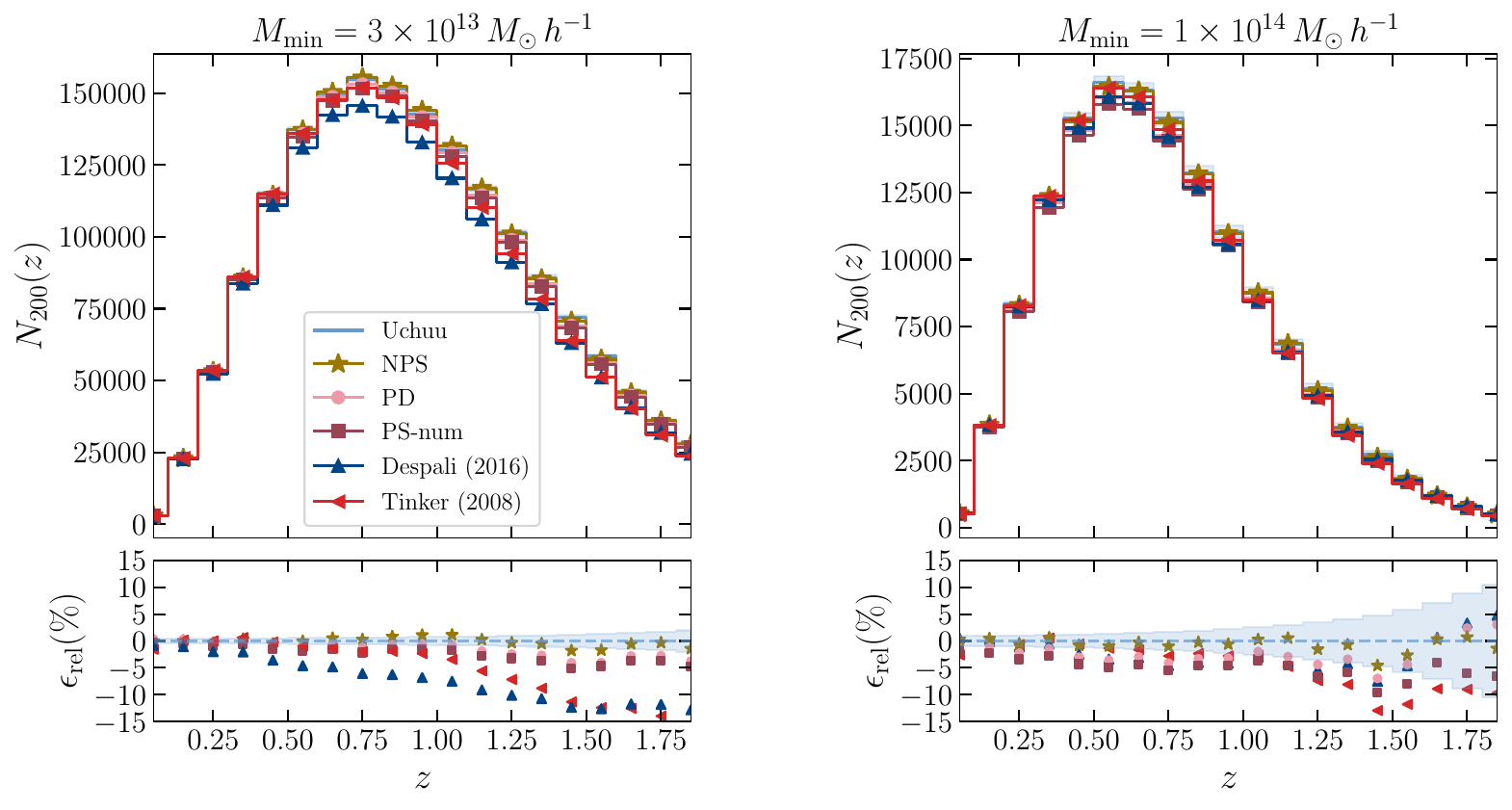}
\includegraphics[width = 0.9\linewidth]{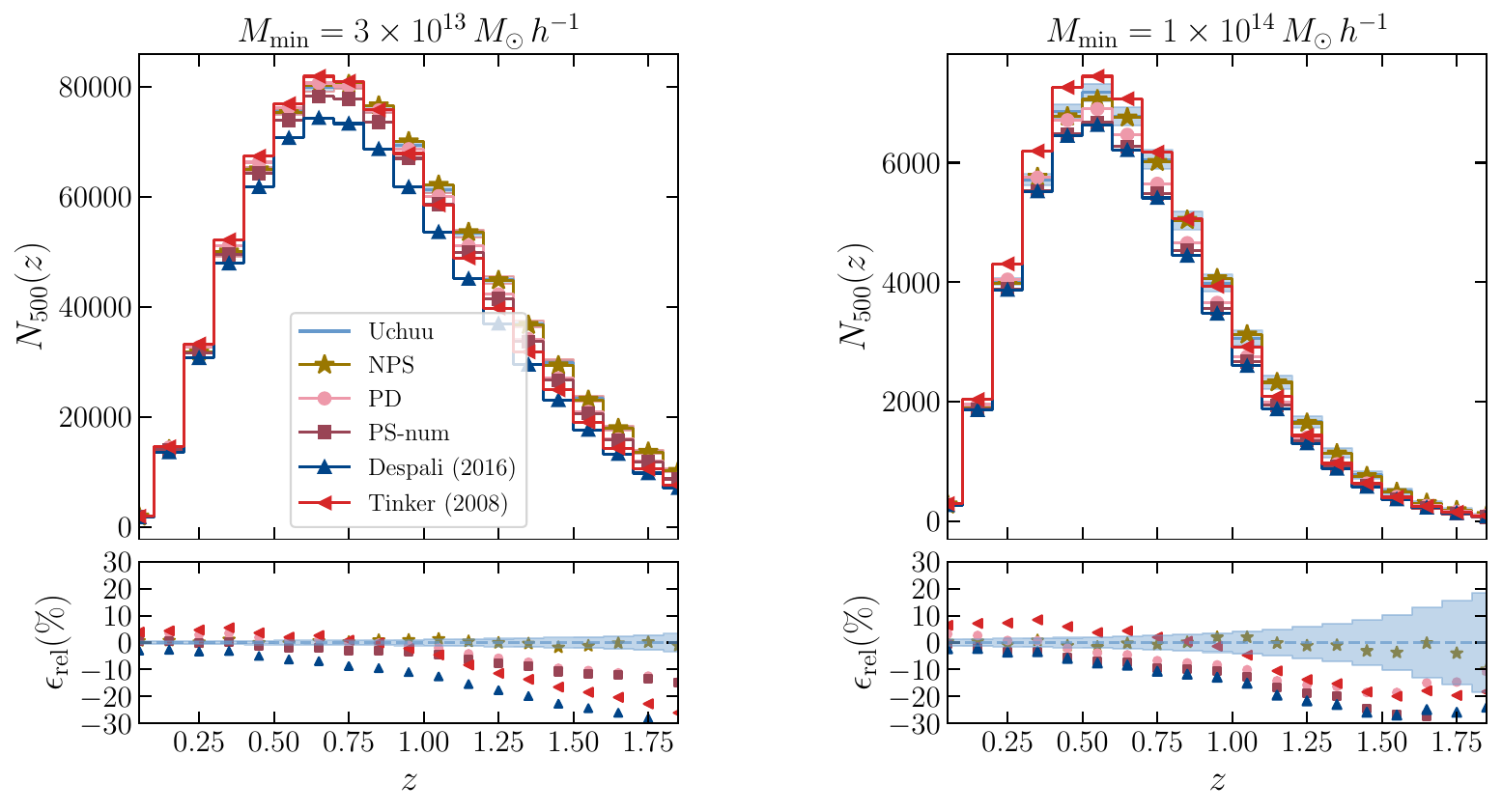}
\caption{\textit{Top panels}: Number counts as function of redshift obtained from the Uchuu-HMF at $M_{200}$ (solid light blue line) for haloes with masses $M_{200}\geq 3\times 10^{13}\,M_{\odot}\,h^{-1}$ ({\it left panel}) and $M_{200}\geq 1\times 10^{14}\,M_{\odot}\,h^{-1}$ ({\it right panel}) in redshift bins of size $\Delta{z}=0.1$ in the case a survey with sky coverage of $15\,000$ deg$^2$. The other curves correspond to the number counts obtained assuming the {\it Euclid}\textbackslash Uchuu-HMF for the different mass conversion models: non-parametric stochastic (solid goldenrod line with star marker), parametric deterministic (solid pink line with circle marker), parametric stochastic performed assuming the numerical distributions of the Uchuu halo concentrations (solid brown line with square marker) and second, the predictions obtained assuming  \citetalias{2016MNRAS.456.2486D} (solid dark blue line with triangle marker) and \citetalias{Tinker_2008} (solid red line triangle-left marker). The lower plots in each panel show the relative difference with respect to the Uchuu data, where the shaded area corresponds to the Poisson errors. 
\textit{Bottom panels}: As in the {\it top panels}, but for haloes with masses $M_{500}$ in the case of the low-mass cut sample with $M_{500}\geq 3\times 10^{13}\,M_{\odot}\,h^{-1}$ (left panel) and high-mass cut, with $M_{500}\geq 10^{14}\,M_{\odot}\,h^{-1}$ (right).}
\label{fig:n200c_combined}
\end{figure*}

\subsection{Number counts from mass converted HMFs}
In Fig.~\ref{fig:n200c_combined}, we plot the number counts 
from the Uchuu data against those obtained using the {\it Euclid}\textbackslash Uchuu-HMF parametrisation at $M_{\rm vir}$ and converted to $M_{200}$ (top panels) and $M_{500}$ (bottom panels) for the low-mass cut, $M_{\rm min}= 3\times 10^{13}$ M$_{\odot}\,h^{-1}$ (left panels) and the high-mass cut $M_{\rm min}= 10^{14}$ M$_{\odot}\,h^{-1}$ (right panels) in the case of the NPS (solid goldenrod lines with star marker), PD (solid pink lines with circle marker), and PS-num (solid brown lines with square marker) mass conversion models. We also plot for comparison the number counts predicted using the \citetalias{Tinker_2008} (solid red lines with triangle-left marker) and \citetalias{2016MNRAS.456.2486D} (solid dark blue lines with triangle marker) HMF parametrisations, which provide fitting functions directly calibrated on halo masses defined at $\Delta=200$ and $500$, respectively. The bottom panels show the relative difference with respect to the Uchuu data along with the associated errors (shaded blue area). The latter were computed by propagating the Poisson errors on the HMF as given by Eq.~(\ref{shot_noise}).

First of all, we may notice that the NPS mass conversion gives number count predictions which are consistent with those from the Uchuu simulation, well within Poisson errors for both mass definitions and mass cuts over the entire redshift interval considered. In the case of the mass conversion to $\Delta=200$, the number counts predicted by the HMFs from \citetalias{Tinker_2008} and \citetalias{2016MNRAS.456.2486D} show the largest discrepancies with respect to the simulation results for both mass cuts, while the PD and PS-num mass conversion result in systematic differences that are of the order of a $1\mbox{--}5\%$ ($2\mbox{--}5\%$) level for the low-mass (high-mass) cut at redshifts of $z\gtrsim 0.2$. We find a similar trend in the case of the mass conversion to $\Delta=500$. Again, the predictions from the \citetalias{Tinker_2008} and \citetalias{2016MNRAS.456.2486D} HMF parametrisation show the largest discrepancies. These systematic deviations will inevitably result in errors in the inferred cosmological parameter constraints. Hence, these results suggest that the use of the \citetalias{Tinker_2008} and \citetalias{2016MNRAS.456.2486D} HMF parametrisations may induce systematic uncertainties on the cosmological parameter constraints that are larger than those induced by the mass conversion models applied to a universally calibrated HMF, which we evaluate next.

\subsection{Systematic effects on cosmological data analyses}
Here, we evaluate the impact of the mass conversion model assumptions on the cosmological parameter constraints from cluster number counts. To this purpose, we assumed the Uchuu counts obtained from the Uchuu-HMF at $M_{200}$ and $M_{500}$ to be our synthetic dataset. Then, we performed a Bayesian parameter inference analysis using the {\it Euclid}\textbackslash Uchuu-HMF at $M_{\rm vir}$ converted to $M_{200}$ and $M_{500}$ to predict the corresponding number counts and infer the cosmological parameter constraints for the different mass conversion approaches. In the following, we exploit the universality of the {\it Euclid}\textbackslash Uchuu-HMF by setting the shape parameters in Eq.~(\ref{nufnu_bhattacharya}) to the values given in Table~\ref{tab:limits_re-calibration}, while the cosmological parameters which specify $\rho_{\rm m}$, $\nu$, ${\rm d}\ln{\nu}/{\rm d}\ln{M}$ in Eq.~(\ref{hmf_formula}) as well as the volume in Eq.~(\ref{volume_element}) are left free to vary.

\subsubsection{Data likelihood}
We assumed a Gaussian likelihood for the number count estimates with errors given by the shot noise of the Uchuu synthetic data. This reads as
\begin{equation}
\mathcal{L}(N_{i}|{\bf
\Theta})=\prod_{i=1}^{N_{i}}\frac{1}{\sqrt{2\pi \sigma_i^2}}\exp\left\{{-\frac{\left[N_{i}-N_{\rm HMF}(z_i|{\bf \Theta})\right]^2}{2\sigma_i^2}}\right\}\;,
\end{equation}
where $N_{i}$ is the synthetic number count estimate in $i$-th redshift bin, $\sigma_i$ is the associated Poisson error, and $N_{\rm HMF}(z_i|{\bf \Theta})$ is the number counts predicted in the $i$-th redshift bin using Eq.~(\ref{cluster_count}) for a given set of cosmological parameters, ${\bf \Theta}$, using a mass converted HMF parametrisation. To sample the parameter space, we adopted a Markov chain Monte Carlo (MCMC) approach using the Metropolis--Hastings algorithm \citep{PhysRevD.66.103511,PhysRevD.87.103529}, as implemented in the publicly available Bayesian analysis framework \texttt{COBAYA} \citep{2021JCAP...05..057T}. We analysed the MCMC chains and plotted the results using the \texttt{GetDist} package \citep{Lewis:2019xzd}. 

In this analysis, we varied five cosmological parameters: $\Omega_{\rm m}$, $\sigma_8$, $H_0$, $\Omega_{\rm b}$, and $n_{\rm s}$. We refer to the review article by \citet{Allen_2011} and references therein for discussions on the cosmological parameter dependence of cluster number counts. Here, it is worth remarking that the abundance of clusters is primarily sensitive to a degenerate combination of $\Omega_{\rm m}$ and $\sigma_8$, which, to zero order, sets the amplitude of Eq.~(\ref{hmf_formula}). Henceforth, we assumed uniform priors for these parameters: $U(0.27,0.33)$ for $\Omega_{\rm m}$, and $U(0.7,1.0)$ for $\sigma_8$. We find the results to be independent of this choice. This is because the large size of the fiducial sample results in small statistical uncertainties, which narrow the MCMC chains around the fiducial cosmology. In contrast, the dependence of the expected number counts on $H_0$, $\Omega_{\rm b}$, and $n_{\rm s}$ is subdominant since they only contribute to the shape of the variance of the matter density field in Eq.~(\ref{variance_HMF}) and the volume element. Consequently, following standard number counts data analyses, we assumed Gaussian priors on these parameters \citep[see e.g.][]{2016A&A...594A..24P}. In particular, we chose their central values to the Uchuu's fiducial cosmology, and assumed a $5\sigma$ standard deviation computed using the values of Table 2, column TT,TE,EE+lowE+lensing from \citep{2020A&A...641A...6P} as adopted in \citep{2022A&A...659A..88L}. The latter prevent to explore unphysical parameter values, which would strongly bias the results. Specifically, we assumed $\Omega_{\rm b}$ to be drawn from the prior distribution ${\cal N}(0.0486,0.0017)$\footnote{${\cal N}(\mu,\sigma)$ denotes a normal distribution with mean $\mu$ and standard deviation $\sigma$.}, and $n_{\rm s}$, from ${\cal N}(0.967, 0.021)$. For $H_0$, we adopt $H_0[\rm km\,s^{-1}\,Mpc^{-1}]$ being drawn from ${\cal N}(67.74,1.00)$. Assuming flat or larger priors would only emphasise the fact that $H_0$, $\Omega_{\rm b}$, and $n_{\rm s}$ are poorly constrained by cluster counts only \citep[see e.g.][]{2021A&A...649A..47A}.

\begin{figure*}[ht]
\centering
\includegraphics[width = 0.49\linewidth]{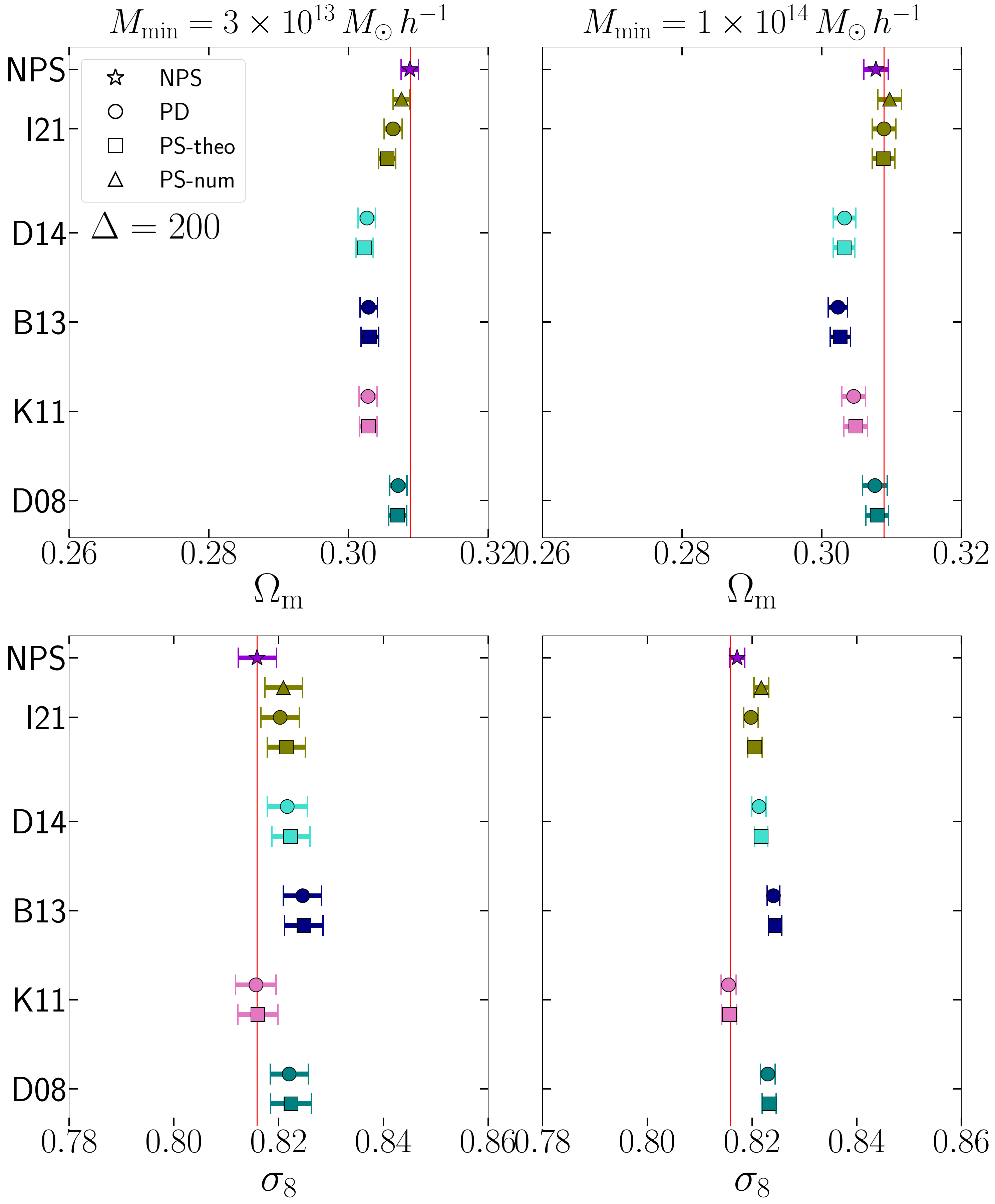}
\includegraphics[width = 0.49\linewidth]{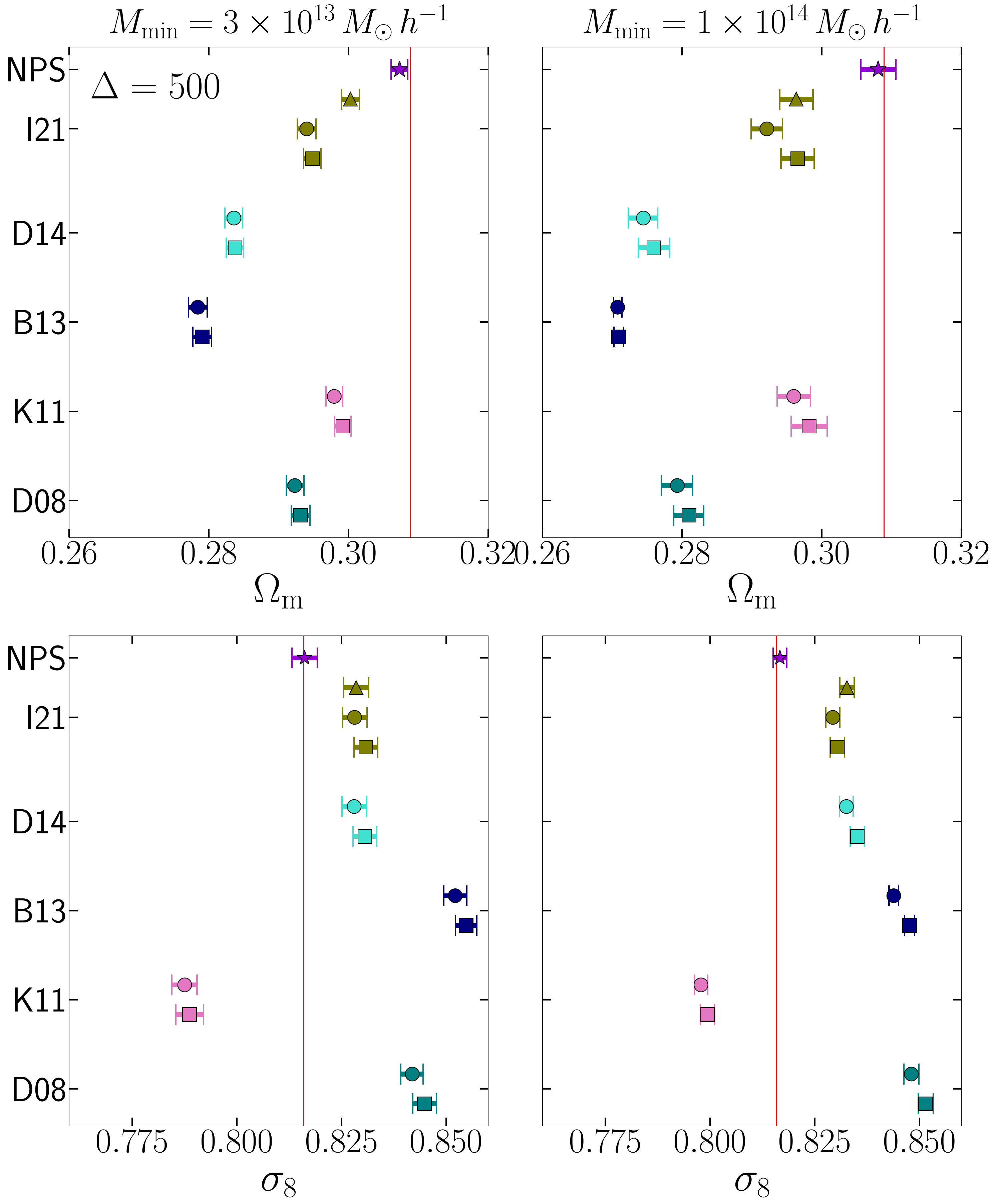}
\caption{Mean and standard deviation of the marginalised constraints on $\Omega_{\rm m}$ ({\it top panels}) and $\sigma_8$ ({\it bottom panels}) inferred from the analysis of Uchuu data at $M_{200}$ ({\it left panels}) and $M_{500}$ ({\it right panels}). In each panel, the left-hand (right-hand) side corresponds to the low (high) -mass cuts. These were obtained by applying the mass conversion to the {\it Euclid}\textbackslash Uchuu-HMF calibrated at $M_{\rm vir}$ using the NPS (magenta star points), the PD (filled circles), the PS-num (filled triangles), and the PS-theo (filled squares) mass conversion approaches. In the PD and PS-theo cases, we have assumed concentration-mass relation from the Uchuu dataset \citetalias{Ishiyama_2021} (olive green) and the relations from \citetalias{2014MNRAS.441.3359D} (cyan), \citetalias{2013ApJ...766...32B} (dark blue), \citetalias{2011ApJ...740..102K} (violet), and \citetalias{2008MNRAS.390L..64D} (dark green). The vertical lines show the fiducial values of $\Omega_{\rm m}$ and $\sigma_8$, respectively.}
\label{fig:summary_plot_cM_omegam}
\end{figure*}

\subsubsection{Cosmological parameter constraints}\label{sec:inference_Uchuu}
The results of the MCMC likelihood data analysis are summarised in Fig.~\ref{fig:summary_plot_cM_omegam}. In particular, the plots show the marginalised mean and standard deviation of $\Omega_{\rm m}$ (top panels) and $\sigma_8$ (bottom panels) obtained from the Uchuu number counts at $\Delta=200$ (left panels) and $\Delta=500$ (right panels). In each panel, the left-hand (right-hand) side shows the results for the low (high) -mass cuts. 

We find that the NPS approach always recovers the Uchuu fiducial cosmology within $1\sigma$ for both $\Delta=200$ and $500$. In the case of $\Delta=200$, the PS-num approach recovers the fiducial value of $\Omega_{\rm m}$ within $1\sigma$ and that of $\sigma_8$ within $2\sigma$, while the PS-theo and PD methods, which assume the median concentration-mass relation from the Uchuu catalogues \citetalias{Ishiyama_2021}, recover the fiducial cosmological parameter values within $2\sigma$. Instead, in the case of the mass conversion to $M_{500}$, the fiducial values are excluded at more than $2\sigma$. This is a direct consequence of the fact that the conditional distribution of the NFW-sparsities maximally differs from that of the true sparsities in the case $s_{{\rm vir},500}$.

It is important to stress that these results have been inferred for an idealised scenario, since the mass conversion models have been calibrated using the conditional sparsity distributions obtained from the Uchuu simulation, which also provides the synthetic data samples at $M_{200}$ and $M_{500}$. Moreover, we have inferred constraints assuming only Poisson errors. Hence, the fact that even in such an idealised case the NFW-based approaches do not recover the fiducial cosmology indicates that the use of the PS and PD mass conversion can introduce a systematic bias in the cosmological parameter inference analysis, which is not the case with the NPS approach.

In practice, as it can be deduced from the plots shown in Fig.~\ref{fig:n200c_combined}, one cannot exclude that larger statistical errors on the cluster number counts, or restricting the analysis to redshift intervals where the number counts predicted by the PD and PS mass conversion models deviates the least from those of the Uchuu fiducial cosmology, would reabsorb such systematic effect within the marginalised $1\sigma$ errors. Moreover, it is well known that cluster number counts probe a degenerate combination of $\Omega_{\rm m}$ and $\sigma_8$. In such a case, it may occur that the bias is attenuated along the parameter degeneracy. Although we find that this is not the case for $S_8=\sigma_8\sqrt{\Omega_{\rm m}/0.3}$, and we refer the reader to Appendix~\ref{S8_degeneracy} for a more detailed discussion.  

Conversely, it is also possible that the deviations in the predicted number counts can be compensated by assuming a different parametrisation of the concentration-mass relation. To check for this eventuality, we have performed additional MCMC likelihood analyses assuming the concentration-mass relation by \citet[][hereafter \citetalias{2008MNRAS.390L..64D}]{2008MNRAS.390L..64D}, \citet[][hereafter \citetalias{2011ApJ...740..102K}]{2011ApJ...740..102K}, \citet[][hereafter \citetalias{2013ApJ...766...32B}]{2013ApJ...766...32B}, and \citet[][hereafter \citetalias{2014MNRAS.441.3359D}]{2014MNRAS.441.3359D}, which are shown in Fig.~\ref{fig:summary_plot_cM_omegam}. We can see that in the case of $M_{500}$, none of the $c$--$M$ relations recovers the fiducial cosmology. In the case of $M_{200}$, the level of bias depends on the mass cut and the cosmological parameter considered. As an example, the use of the $c$--$M$ relation from \citetalias{2011ApJ...740..102K} recovers the fiducial value of $\sigma_8$ within $1\sigma$ for both mass cuts, while it does not recover the value of $\Omega_{\rm m}$ for both mass cuts. In contrast, the use of $c$--$M$ relation from \citetalias{2013ApJ...766...32B} leads to biased results at more than $1\sigma$ for both parameters.

Overall, the analysis presented here indicates that the adoption of the NPS mass conversion does not introduce systematic effects larger than the level of Poisson noise. This is not the case with the PS and PD methods. The adoption of different $c$--$M$ relations can mitigate these effects only for mass conversion to $M_{200}$, since it is only for such a mass definition that the conditional NFW-sparsity distribution differs the least from that of the true sparsities. However, this may depend on the specific cosmological model underlying the number count data. This cannot be known a priori; thus, if the PS or PD have to be used, one should always test the stability of the results under different $c$--$M$ relations.

\begin{figure}[ht]
\centering
\includegraphics[width = 0.9\linewidth]{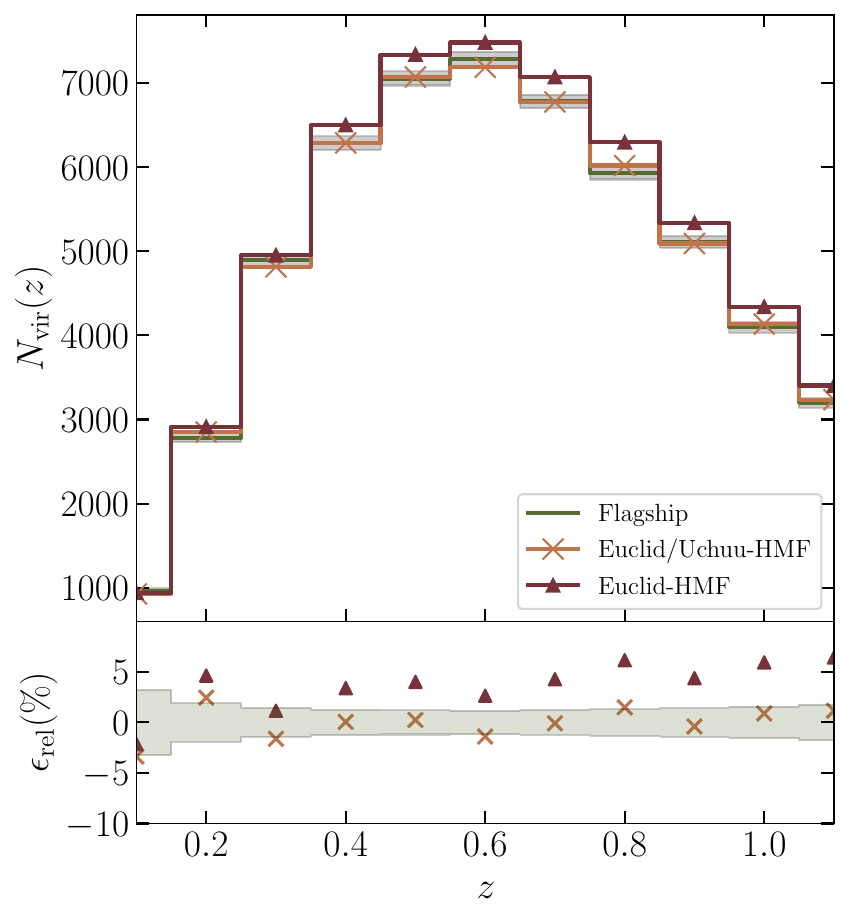}
\caption{{\it Top panel}: Number counts of haloes detected at $M_{\rm vir}$ from the Flagship light-cone catalogue (solid green line) with $M_{\rm vir}\ge 10^{14}\,M_{\odot}h^{-1}$ over a sky area of ${\pi}/{2}$ \si{\steradian} in bins of size $\Delta{z}=0.1$ against the predictions from the {\it Euclid}\textbackslash Uchuu-HMF (solid orange line) and the {\it Euclid}-HMF (solid brown line). {\it Bottom panel}: Relative difference with respect to the Flagship number counts. The shaded area corresponds to the Poisson noise.}
\label{fig:flagship_ncount_mvir}
\end{figure}

\section{Cosmological analysis of Flagship cluster counts}\label{sec:mass_conv_flag}
In the previous section, we have evaluated the impact of the different mass conversion models on the cosmological parameter inference analysis of cluster number counts for an idealised case. Here, we investigate these effects using synthetic number counts generated from a simulation that differs from that used to calibrate the HMF at $M_{\rm vir}$, as well as the estimation of the conditional sparsity distributions. 
\subsection{Flagship synthetic samples and likelihood}

To this purpose, we used the WIDE light-cone halo catalogue from the Flagship simulation described in Sect.~\ref{sec:flagship}, which provides an approximation of the halo population to be detected by \textit{Euclid} observations. However, since the Flagship simulation has a higher mass resolution than the PICCOLO suite used in the calibration of the {\it Euclid}-HMF, we first test for any systematic difference between the predicted number counts at $M_{\rm vir}$ and the one from the Flagship light-cone dataset. These are shown in bins of size $\Delta{z}=0.1$ in Fig.~\ref{fig:flagship_ncount_mvir} for the {\it Euclid}-HMF and {\it Euclid}\textbackslash Uchuu-HMF fitting functions. We can see that the {\it Euclid}-HMF tends to over-predict the number counts by $3-6\%$ for $0.2\lesssim z\lesssim 1$.  Hence, we may expect differences between the constraints inferred using the two HMFs, and any effect due to the mass conversion approach will compound with such intrinsic differences. For this reason, we will show results inferred using both HMFs. Given the consistency of the PS-theo and PS-num results derived from the Uchuu counts in Sect.~\ref{sec:inference_Uchuu}, we restrict the presentation of the Flagship analysis to the PS-theo method, which we hereafter refer to as the PS approach.

We selected two distinct synthetic halo samples, one with mass $M_{200}\geq 10^{14}\,M_{\odot}\,h^{-1}$ and one with $M_{500}\geq 10^{14}\,M_{\odot}\,h^{-1}$. We binned these samples in bins of size $\Delta{z}=0.1$ over the redshift interval $0.1\le z\le z_{\rm max}$ and considered two distinct cases with $z_{\rm max}=0.8$ and $1.2$. Since the comoving distance with respect to the origin is given for all haloes in the light-cone catalogue, we evaluated the comoving distance corresponding to the edge of the redshift bins and counted the number of haloes inside the distance range of each redshift bin. We compared these synthetic data to predictions obtained using Eq.~(\ref{cluster_count}) for the {\it Euclid}-HMF and the {\it Euclid}\textbackslash Uchuu-HMF mapped to the overdensity of interest. In evaluating Eq.~(\ref{cluster_count}), we have checked that errors due to the small redshift bin size approximation are smaller than $1\%$.

We assumed a Gaussian likelihood for the Flagship data number counts,
\begin{equation}
\mathcal{L} = \frac{\exp\left\{-\frac{1}{2}\left[\mathbf{N}-\mathbf{N_{\rm HMF}}({\bf \Theta})\right]^{\rm T} \mathbf{C}_\Delta^{-1}\left[\mathbf{N}-\mathbf{N_{\rm HMF}}({\bf \Theta})\right]\right\}}{\sqrt{(2\pi)^n\,\textrm{det}\,{C_{\Delta}}}},
    \label{eq:likelihood_flagship}
\end{equation}
where $\rm \mathbf{N}$ is the data vector of dimension $n$ containing the number counts in each redshift, $\mathbf{N}_{\rm HMF}$ is the theoretical model prediction for a given set of cosmological parameters, ${\bf \Theta}$, using a given halo mass converted HMF, and $\mathbf{C}_\Delta$ is the data covariance matrix. We refer the readers to Appendix~\ref{data_cov} for a detailed description of the computation of the covariance. Similarly to the analysis of the synthetic Uchuu dataset, we sampled the posterior distribution by adopting an MCMC approach using the \texttt{COBAYA} implementation of the Metropolis--Hastings algorithm. From this posterior we inferred constraints on five cosmological parameters: $\Omega_{\rm m}$, $\sigma_8$, $H_0$, $\Omega_{\rm b}$, and $n_{\rm s}$. Again, we assumed Gaussian priors on $\Omega_{\rm b}$, $H_0[{\rm km\,s^{-1}\,Mpc^{-1}}]$ and $n_{\rm s}$, namely ${\cal N}(0.049,0.001)$, ${\cal N}(67,1)$ and ${\cal N}(0.96,0.02)$, respectively, and uniform priors on $\Omega_{\rm m}$ and $\sigma_8$, given by $U(0.27,0.33)$ and $U(0.7,1.0)$, respectively.

Here, it is worth stressing that Flagship and Uchuu fiducial cosmologies have values of $S_{8}=\sigma_8\sqrt{\Omega_{\rm m}/0.3}$ that differ at the $~1\%$ level. Since this is the parameter to which the sparsity is most sensitive, the small difference implies that we can use the conditional halo sparsity distributions calibrated using the Uchuu simulation. For a more general treatment, where the fiducial cosmology of the synthetic data is unknown, one needs to handle the cosmology dependence of the sparsity by using, for instance, an emulator (S\'aez-Casares et al., in prep.).

\subsection{Cosmological parameter constraints}
\begin{figure*}[th]
\centering
\includegraphics[width = 0.49\linewidth]{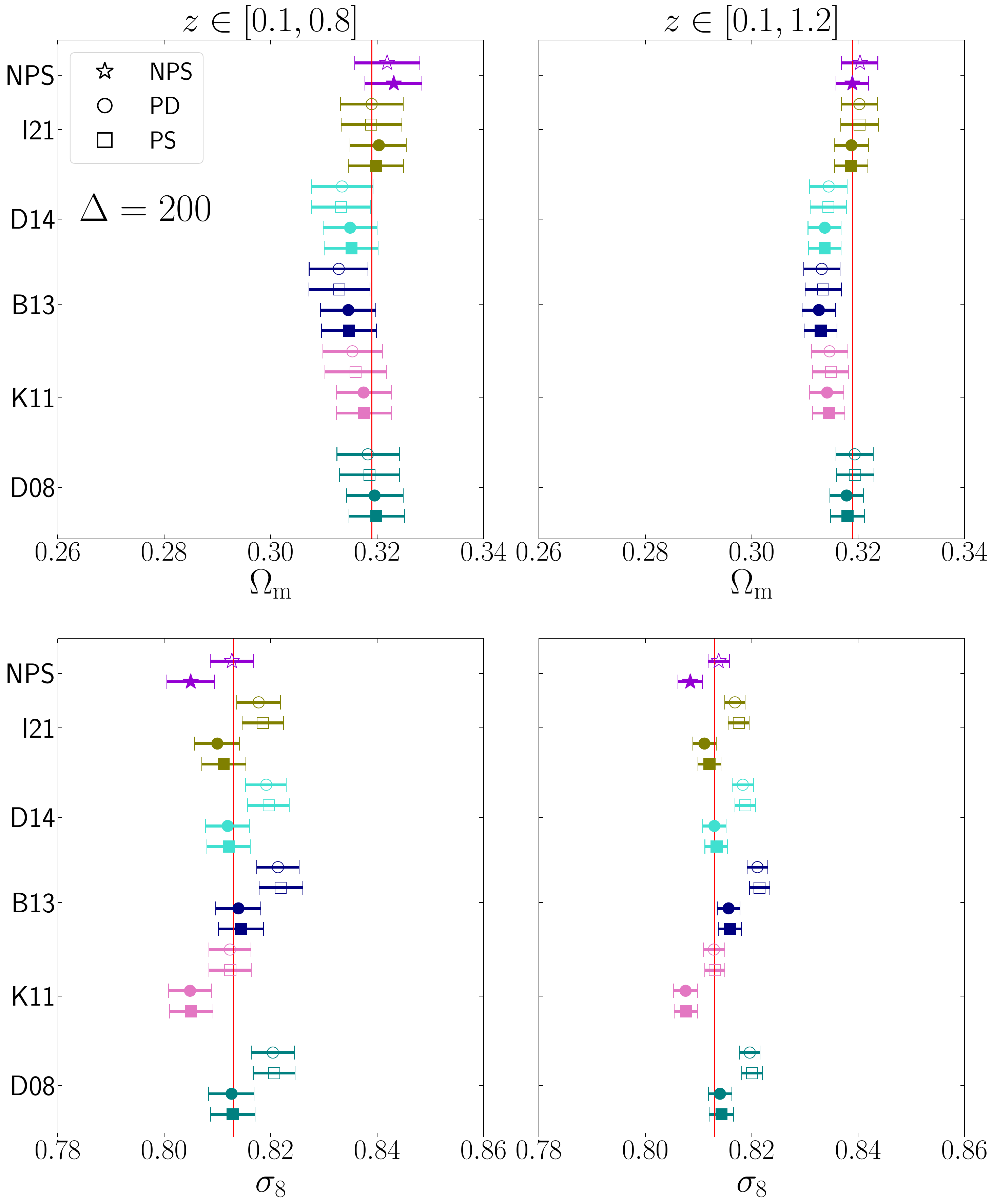}
\includegraphics[width = 0.49\linewidth]{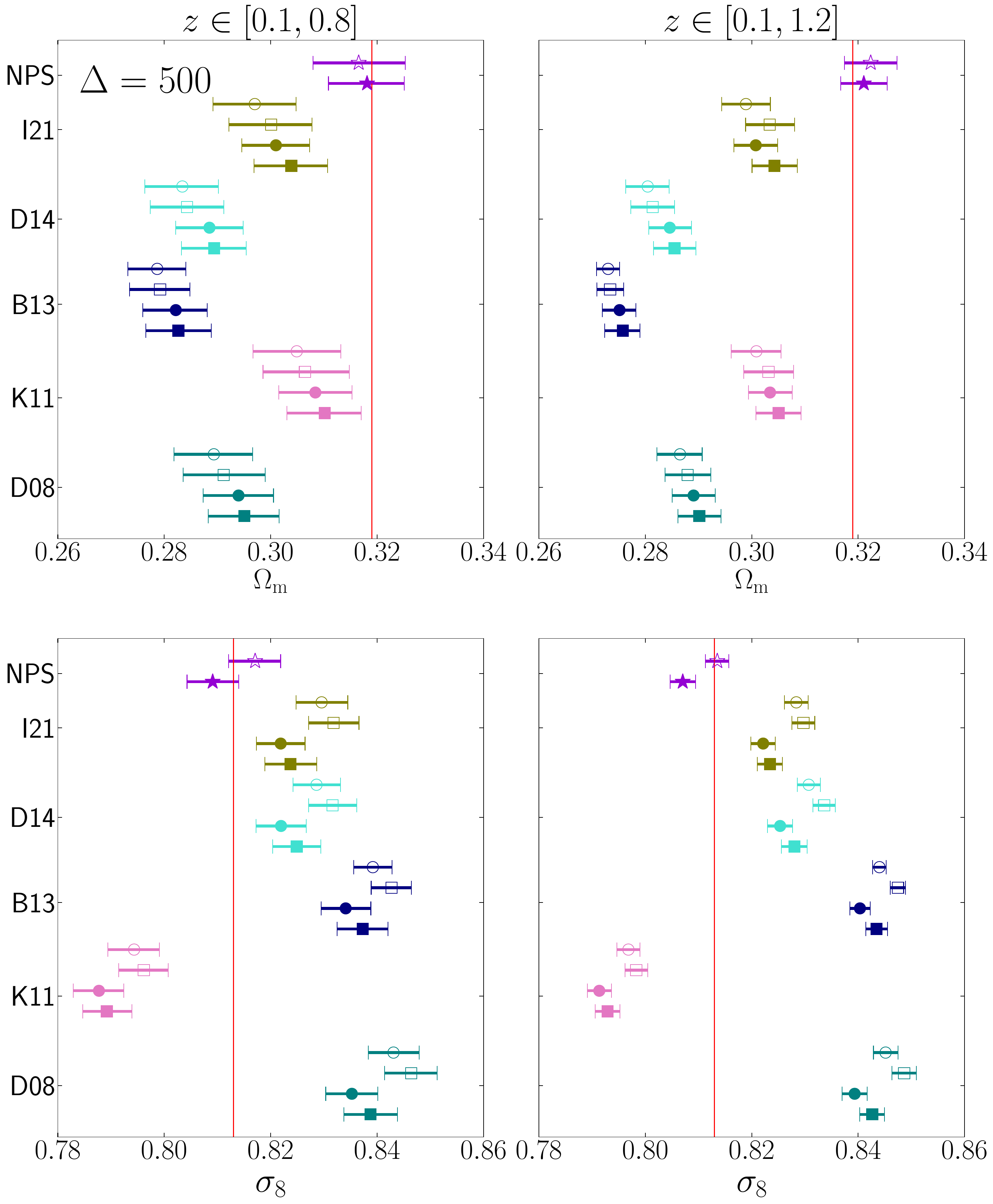}
\caption{Mean and standard deviation of the marginalised constraints on $\Omega_{\rm m}$ ({\it top panels}) and $\sigma_8$ ({\it bottom panels}) inferred from the analysis of Flagship data at $M_{200}$ ({\it left panels}) and $M_{500}$ ({\it right panels}). In each panel, the left-hand (right-hand) side corresponds to the low (high) redshift cut. These have been obtained using the NPS mass conversion (magenta stars), the PD (circles), and PS (squares) approaches for different concentration-mass relations as in Fig.~\ref{fig:summary_plot_cM_omegam}. The empty (filled) symbols correspond to the constraints inferred assuming the {\it Euclid}\textbackslash Uchuu-HMF ({\it Euclid}-HMF) mapped to the target mass definition of the Flagship datasets. The vertical lines show the fiducial value of $\Omega_{\rm m}$ and $\sigma_8$,  respectively.}
\label{fig:summary_plot_cM_omegam_flagship}
\end{figure*}

In Fig.~\ref{fig:summary_plot_cM_omegam_flagship}, we plot the marginalised constraints on $\Omega_{\rm m}$ and $\sigma_8$ inferred from the analysis of the synthetic Flagship number counts for different mass definitions and model assumptions. First, we can see that, as in the case of the Uchuu data analysis, the {\it Euclid}\textbackslash Uchuu-HMF with the NPS method (empty magenta stars in Fig.~\ref{fig:summary_plot_cM_omegam_flagship}) always recovers the fiducial cosmological parameter values within a level of $1\sigma$. Comparing these constraints to those obtained from the analysis of the Uchuu number counts with high-mass cut (filled magenta stars in Fig.~\ref{fig:summary_plot_cM_omegam}) and discussed in Sect.~\ref{sec:inference_Uchuu}, we find that the deviations with respect to the fiducial cosmological parameter values follow a similar pattern for both $\Delta=200$ and $500$. This is also the case for the constraints inferred assuming the PS and PD approaches, assuming different concentration-mass relations. In all these cases, the only noticeable difference between the constraints inferred from the Uchuu and Flagship samples is the fact that the latter results in a larger standard deviation of the constrained parameters, which is consistent with the different level of Poisson noise associated with Flagship and Uchuu synthetic datasets. This is an important consistency check given that the two simulations have nearly identical fiducial cosmological parameter values. 

Let us now focus on the results of the Flagship data analysis obtained under the {\it Euclid}-HMF for the NPS mass conversion (filled magenta stars in Fig.~\ref{fig:summary_plot_cM_omegam_flagship}). For both cases, $\Delta=200$ and $500$, the fiducial value of $\Omega_{\rm m}$ is recovered within $1\sigma$ for both redshift cuts, while the value of $\sigma_8$ is recovered within $1\sigma$ ($2\sigma$) for the low (high) redshift cut. Also, we may notice that the mean value of $\sigma_8$ is systematically smaller than that inferred assuming the {\it Euclid}\textbackslash Uchuu-HMF. Hence, depending on the specific dataset, the adoption of the {\it Euclid}-HMF may result in slightly biased constraints compared to the {\it Euclid}\textbackslash Uchuu-HMF. As shown in Fig.~\ref{fig:flagship_ncount_mvir}, this is a direct consequence of the fact that the {\it Euclid}-HMF tends to over-predict the simulation number counts at a $\sim 5\%$ level in the range $0.2\lesssim z\lesssim 1.0$, and thus requires lower values of $\sigma_8$ to match the Flagship data. 

Having presented the results of the NPS mass conversion, let us discuss the constraints inferred from the PS and PD mass conversions. As we can see in Fig.~\ref{fig:summary_plot_cM_omegam_flagship}, the ability of these approaches to recover the fiducial cosmology depends on the target mass definition, the assumed HMF, the assumed concentration-mass relation, as well as the redshift range of the synthetic data sample.  In particular, in the case of the mapping to $\Delta=200$, we can see that the different $c$--$M$ relations constrain the value of $\Omega_{\rm m}$ within $1\sigma$ of the fiducial value for the low redshift cut, while in the case of the higher redshift cut, this depends on the assumed $c$--$M$ relation. For $\sigma_8$, we find that the ability to recover its fiducial value depends on the assumed HMF for both redshift samples. Again, this is because of the intrinsic differences in the {\it Euclid}-HMF and {\it Euclid}\textbackslash Uchuu-HMF in reproducing the Flagship number counts, which can be compensated by the effects induced in the number count predictions by the redshift and mass dependence of the assumed concentration-mass relation. Furthermore, in the case of $\Delta=500$, we can see that the use of PD and PS mass conversion can introduce a systematic bias at high statistical significance. We address the reader to Appendix~\ref{S8_degeneracy_flagship} for a discussion on the impact of the halo mass conversion on the $S_8$ parameter.

Overall, this analysis suggests that the use of the NPS mass conversion does not introduce statistically significant systematic errors on the cosmological parameter constraints. In the case of the PS and PD methods, an unbiased value of $\Omega_{\rm m}$ is recovered only for the mass conversion to $M_{200}$, while for $\sigma_8$ this depends on the assumed HMF and concentration mass relation. In contrast, the PS and PD approaches can induce a large bias effect in the case of the mass conversion to $M_{500}$, whose amplitude varies with the choice of the assumed concentration-mass relation. However, this may depend on the specific cosmological model underlying the number count data. As such, if the PS or PD methods are used in observational data analyses, it is important to gauge the stability of the results under different concentration-mass relations.

\section{Conclusions}\label{sec:conclu}
Cosmological analyses of the {\it Euclid} cluster number counts will rely on the adoption of an accurate parametrisation of the HMF capable of capturing the cosmological parameter dependence of the abundance of clusters \citep{Castro-EP24}. 
These will require the use of scaling mass-observable relations also calibrated at $\Delta_{\rm vir}$. Alternatively, to accurately predict the observed cluster counts, the adoption of scaling relations calibrated at different overdensities, such as $\Delta = 200$ or $500$, will require a map of the HMF to the observed mass definition.

Here, we have investigated the impact of different mass conversion methods on the cosmological parameter constraints from cluster number count data analysis using the general formalism introduced by \citet{2023A&A...674A.173R}. This makes use of the halo sparsity statistics, which naturally emerges as an essential ingredient to correctly map the HMF at two different mass definitions, providing a non-parametric stochastic (NPS) approach to perform the HMF mass conversion without the need to assume a specific form of the halo density profile. The same formalism can also integrate the standard mass conversion approach based on assuming the NFW profile specified by a given concentration-mass relation, which we referred to as the parametric deterministic (PD) method. Similarly, it also makes it possible to account for the scatter of the concentration parameter, which we have referred to as the parametric stochastic (PS) approach. 

We have derived cosmological parameter constraints from synthetic datasets to evaluate any systematic bias induced by these different mass conversion approaches on cluster number count analyses. For these purposes, we have used halo catalogues from the Uchuu and Flagship simulations. We find that the NPS approach always recovers the fiducial cosmology independently of the observational configuration considered, while the PS and PD methods can introduce a significant source of bias, depending on the adopted HMF and the assumed concentration-mass relation. As an example, the analysis of the synthetic Flagship clusters with mass defined at $\Delta=500$ over the redshift range $0\le z\le 1.2$ indicates that the PS and PD methods are unable to recover the fiducial values of $\Omega_{\rm m}$ and $\sigma_8$ at more than $2\sigma$. Hence, in the case where the PS and PD approaches are used, we advocate for testing the stability of the results assuming different concentration-mass relations.

We would like to point out that this is the first analysis dedicated to the impact of the halo mass conversion assumptions on cluster count data analysis. As such, the work presented here can be extended in several directions. Firstly, the PD and PS approaches can be applied to parametric profiles other than NFW, such as the Einasto profile. This can be done in the random variable formalism, whose extension to other parametric halo density profiles has already been highlighted in \citetalias{2023A&A...674A.173R}. A more realistic cosmological analysis may include the propagation of the mass-cluster observable relation in the evaluation of the cosmological parameter constraints. The effect of baryons should also be taken into account. As shown in \citet{EP-Castro}, this can be implemented as a differential variable transformation to the universal HMF calibrated at $\Delta_{\rm vir}$. The mapping to the observed mass definition can then be performed with the NPS method, taking advantage of the fact that the impact of baryons on the sparsities at small overdensities is smaller than a few percent \citep{2025A&A...697A..33C}. In such a case, it would also be very interesting to compare the approach presented here with that in which the HMF is kept at the virial mass definition, and the mass conversion is applied to the scaling relations. We leave this to future work. 

\begin{acknowledgements}
The authors thank the anonymous referees for their careful reading of the manuscript and the constructive feedback they provided. TG thanks Charlotte Rizo-Crastes for useful discussions on $\texttt{ROCKSTAR}$ halo catalogues' specificities. 
TG, PSC, and AMCLB acknowledge funding APR 4eedb6a7b6 from CNES. LM acknowledges the financial contribution from the PRIN-MUR 2022 20227RNLY3 grant “The concordance cosmological model: stress-tests with galaxy clusters” supported by Next Generation EU and from the grant ASI n. 2024-10-HH.0 “Attività scientifiche per la missione Euclid – fase E”

\AckEC  \AckCosmoHub
We thank Instituto de Astrofisica de Andalucia (IAA-CSIC), Centro de Supercomputacion de Galicia (CESGA), and the Spanish academic and research network (RedIRIS) in Spain for hosting Uchuu DR1, DR2, and DR3 in the Skies and Universes site for cosmological simulations. The Uchuu simulations were carried out on Aterui II supercomputer at Center for Computational Astrophysics, CfCA, of National Astronomical Observatory of Japan, and the K computer at the RIKEN Advanced Institute for Computational Science. The Uchuu Data Releases efforts have made use of the ${\rm skun@IAA_RedIRIS}$, and ${\rm skun6@IAA}$ computer facilities managed by the IAA-CSIC in Spain (MICINN EU-Feder grant EQC2018-004366-P). This work made use of Astropy:\footnote{http://www.astropy.org} a community-developed core Python package and an ecosystem of tools and resources for astronomy \citep{astropy:2013, astropy:2018, astropy:2022}.
\end{acknowledgements}

\bibliography{Euclid}

\begin{appendix}

\section{HMF fitting function and calibration}\label{HMFcalibration}
\subsection{HMF parametrisation}
The number density of haloes with mass in the infinitesimal interval $[M,M+{\rm d}M]$ can be written as \citep{1974ApJ...187..425P,1991ApJ...379..440B,Sheth_2001}
\begin{equation}\label{hmf_formula}
\frac{{\rm d}n}{{\rm d}\ln{M}}=\frac{\rho_{\rm m}}{M}\nu f(\nu)\frac{{\rm d}\ln{\nu}}{{\rm d}\ln{M}}\;,
\end{equation}
where $\rho_{\rm m}$ is the comoving mean matter density and $\nu f(\nu)$ is the multiplicity function, which encodes the effects of the non-linear gravitational processes that determine the formation of haloes. This is usually expressed in terms of the peak height variable $\nu=\delta_{\rm c}/\sigma(M,z)$; that is, the ratio of the linearly extrapolated spherical collapse overdensity threshold at $z=0$ and the root-mean-square deviation of the linear matter density field at a given redshift. In the following, we assumed the fitting formula of \citet{1996ApJ...469..480K},
\begin{equation}
\delta_c(z) = \frac{3}{20}\left(12\pi\right)^{2/3}\left[1+0.0123\log_{10}\Omega_{\rm m}(z)\right]\;,
\end{equation}
and we computed the variance of the linear matter density field within a spherical region of radius $R$, which encloses a mass of $M=(4/3)\pi \rho_{\rm m} R^3$,
\begin{equation}\label{variance_HMF}
\sigma^2(M,z)=\frac{1}{2\pi^2}\int_0^{\infty}k^2P_{\rm m}^{\rm lin}(k,z)\tilde{W}_{R}^2(k)\,{\rm d}k\;,
\end{equation}
where $P_{\rm m}^{\rm lin}(k,z)$ is the linear matter power spectrum at redshift $z$, which we computed using the analytical approximation of the linear transfer function from \citet{1998ApJ...496..605E} and $\tilde{W}_R(k)$ is the Fourier transform of spherical top-hat window function in real space. All these functions were computed using the cosmology Python package \texttt{colossus} \citep{2018ApJS..239...35D}.

\citetalias{Castro-EP24} adopted the parametrisation of the multiplicity function, originally introduced by \citet{Bhattacharya2010MASSFP}, which reads
\begin{equation}\label{nufnu_bhattacharya}
    \nu f(\nu) = A(p,q)\sqrt{\frac{2a\nu^2}{\pi}}e^{-a\nu^2/2}\left(1+\frac{1}{(a\nu^2)^p}\right ) (\nu\sqrt{a})^{q-1}\;,
\end{equation}
where $a$, $p$, and $q$ are promoted to parametrised fitting functions that aim to capture the cosmological dependence of the multiplicity function beyond that encoded in the peak height variable. These are written as  
\begin{align}
    a(z,M) &= a_R(z,M)\,\Omega^{\rm a_z}_{\rm m}(z)\;, \\
    p(z,M) &= p_1 + p_2\left(\frac{{\rm d}\ln\sigma}{{\rm d}\ln R} + 0.5 \right)\;, \\
    q(z,M) &= q_R(z,M)\,\Omega^{\rm q_z}_{\rm m}(z)\;,
\end{align}
with
\begin{align}
    a_R(z,M) &= a_1 + a_2\left(\frac{{\rm d}\ln\sigma}{{\rm d}\ln R} + 0.6125\right)^2\;, \\
    q_R(z,M) &= q_1 + q_2\left(\frac{{\rm d}\ln\sigma}{{\rm d}\ln R} + 0.5\right)\;.
\end{align}
Finally, the normalisation function $A(p,q)$ is given by
\begin{equation}
    A(p,q) = \left\{\frac{2^{-1/2-p+q/2}}{\sqrt{\pi}} \left[2^p\Gamma\left(\frac{q}{2}\right)+\Gamma\left(-p+\frac{q}{2}\right) \right] \right\}^{-1}\;,
\end{equation}
where $\Gamma$ is the Gamma function. In summary, Eq.~(\ref{nufnu_bhattacharya}) depends on the parameters $\{a_1,a_2,a_z,p_1,p_2,q_1,q_2,q_z\}$. These were calibrated in \citetalias{Castro-EP24} using halo catalogues from a suite of $26$ $N$-body simulations of $(2\,h^{-1}\,{\rm Gpc})^3$ volume with $4\times 1280^3$ particles. These cosmological boxes represent distinct realisations of nine cosmological models. More specifically, the authors used virial halo masses $M_{\rm vir}$ defined as the mass within a radius enclosing the virial overdensity $\Delta_{\rm vir}(z)$ as predicted by the spherical collapse model \citep{1998ApJ...495...80B}. We refer the interested reader to \citetalias{Castro-EP24} for a detailed description of the fitting procedure of the {\it Euclid}-HMF. As already mentioned, we refer to this calibrated HMF as the {\it Euclid}-HMF, specified by the values of the fitting parameters obtained from the analysis of the halo catalogues generated by \texttt{ROCKSTAR} halo finder that are quoted in the first line of Table 4 in \citetalias{Castro-EP24}.  
Since we aim to evaluate systematic errors due to the halo mass conversion, we require a HMF fitting function capable of reproducing the numerical estimate of the Uchuu-HMF within the Poisson errors. For this reason, we performed a Bayesian inference analysis of the Uchuu-HMF data at $\Delta_{\rm vir}$, which we describe next.
\subsection{HMF Bayesian parameter inference}
We infer constraints and best-fit values of the multiplicity function parameters $\Theta=\{a_1,a_2,a_z,p_1,p_2,q_1,q_2,q_z\}$ using the numerical estimates of the HMF from the Uchuu halo catalogues at $M_{\rm vir}$ for the 24 distinct redshift snapshots in the range $0\le z \le 2$.

We assumed a Gaussian likelihood function
\begin{equation}\label{likelihood_HMF}
    \ln \mathcal{L}(\mathbf{x} |\mathbf{\mu} ,\mathbf{\sigma}, \Theta) = \prod_{i=1}^{N_z}\prod_{j=1}^{N_M}\frac{\exp \left[-\frac{1}{2}(x_{ij}-\mu_{ij})^2/\sigma^2_{ij}\right]}{\sqrt{2\pi \sigma^2_{ij}}}\;,
\end{equation} 
where $N_z$ is the number of redshift snapshots, $N_M$ is the number of HMF estimates in different mass bins, $x_{ij}$ is the HMF estimate in the $i$-th redshift, and $j$-th mass bins, $\sigma_{ij}$ is the corresponding Poisson error, and $\mu_{ij}$ is the prediction from the HMF fitting function. We assumed flat priors on the fitting parameters and ran an MCMC sampling of the parameter space using the \texttt{emcee} sampler \citep{2013PASP..125..306F}, a Python implementation of the affine-invariant algorithm \citep{64ebe435829647f594fea6d945067389}. Furthermore, we used 64 walkers initialised within a Gaussian sphere centred around the maximum likelihood estimate of Eq.~(\ref{likelihood_HMF}). The posterior distribution was sampled using $12\,000$ steps. We computed the autocorrelation time $\tau$ for each parameter and found a maximum of approximately 124. To ensure convergence and reduce bias from initialisation, we discarded the first 250 ($\simeq 2\tau$) steps of the chain for each walker. Additionally, we thinned the chain by a factor of 60 ($\simeq \tau/2$) to minimise autocorrelation in the sample.
In Table~\ref{tab:limits_re-calibration}, we quote the best-fit parameter values as well as the marginalised mean and $68\%$ credible interval.
\begin{table}[th]
\centering
\small
\caption{Best fit values, marginalised mean, and $1\sigma$ uncertainty of the fitting parameters of {\it Euclid}\textbackslash Uchuu-HMF obtained from the Bayesian inference analysis of the Uchuu-HMF dataset.}
\begin{tabular}{|c|S[table-format=1.4]|S[table-format=1.3(3)]|}
\hline
$\Theta$ & {$\Theta_{\rm best-fit}$} & {$\bar{\Theta}\pm \sigma_{\Theta}$} \\
\hline
$a_1$  & 0.8129  & 0.812 \pm 0.003 \\
$a_2$  & 0.6225  & 0.637 \pm 0.075 \\
$a_z$  & -0.0436 & -0.042 \pm 0.005 \\
$p_1$  & -0.6443 & -0.642 \pm 0.011 \\
$p_2$  & -0.0915  & -0.098 \pm 0.067 \\
$q_1$  & 0.3567  & 0.357 \pm 0.003 \\
$q_2$  & 0.1784  & 0.178 \pm 0.034 \\
$q_z$  & -0.0618  & -0.061 \pm 0.006 \\
\hline
\end{tabular}
\label{tab:limits_re-calibration}
\end{table}
\section{Conditional sparsity distribution}\label{ConditionalSparsityCalibration}

\begin{figure*}[ht]
\centering
\includegraphics[width = 0.9\linewidth]{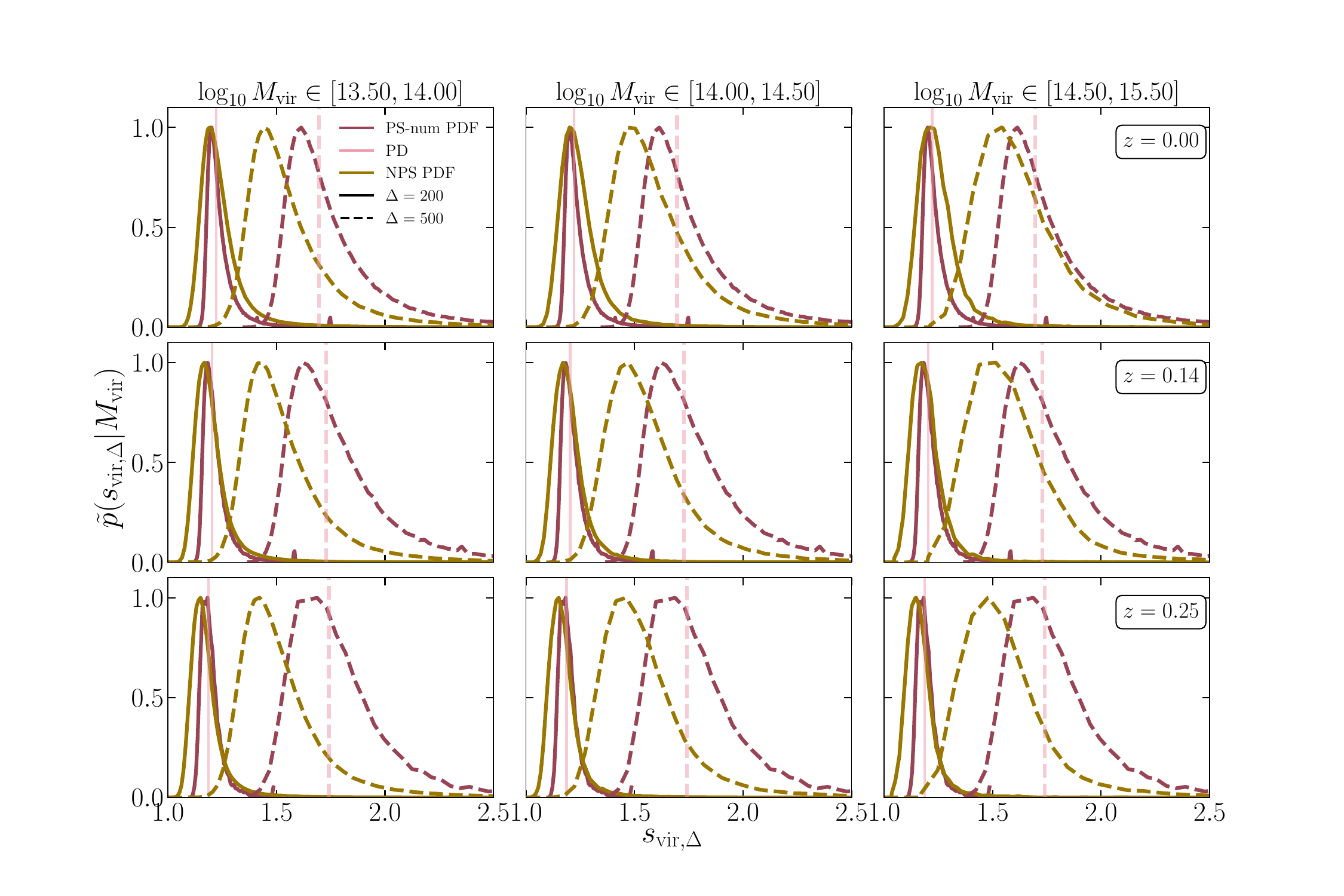}
\caption{Normalised conditional probability density functions of the halo sparsities $s_{{\rm vir},200}$ (solid line) and $s_{{\rm vir},500}$ (dashed line) at $z=0$ ({\it top panels}), $0.14$ ({\it central panels}), and $0.25$ ({\it bottom panels}) for three different bins of mass $M_{\rm vir}\,[M_{\odot}\,h^{-1}]$ ({\it panels from left to right}). In each panel, the curves of different colours correspond to the conditional distribution estimated from the `true' sparsities of the Uchuu haloes (goldenrod curves) defined as the ratio of the halo masses in the Uchuu's catalogue, which is used in the NPS approach; the conditional distributions of the NFW-inferred sparsities (brown curves), i.e. halo sparsities computed using the concentration parameters derived from the Uchuu halo catalog, which are used in the PS-num approach; the NFW-sparsity value given by the Uchuu's concentration-mass relation at the central mass bin (pink vertical line) used in the PD approach. For visual purposes, we have normalised the distributions of the NFW-sparsities to the peak value of the sparsity distributions.}
\label{fig:uchuu_pdfs}
\end{figure*}
In order to perform the halo mass conversion using the NPS approach, we need to estimate the conditional probability density function of the halo sparsities. Since we are interested in mapping the HMF from $M_{\Delta_1}\equiv M_{\rm vir}$ to the HMF at $M_{\Delta_2}\equiv M_{200}$ and $M_{500}$, we computed the sparsities $s_{{\rm vir},200}$ and $s_{{\rm vir},500}$ for every halo in the Uchuu catalogues and then estimated the conditional distributions $p_s(s_{{\rm vir},200}|M_{\rm vir})$ and $p_s(s_{{\rm vir},500}|M_{\rm vir})$. Similarly, in the case of the PS-num approach, we need to estimate the conditional distributions of the halo concentrations. We consider two distinct cases, which we refer to as PS-num and PS-theo, respectively. In the PS-num case, we computed the halo concentration $c_{\rm vir}$ using the scale and virial radius from the Uchuu halo catalogues and converted them into sparsities using the continuous differentiable relation provided in \citetalias{2023A&A...674A.173R}. Finally, we used these numerical NFW-sparsities to compute the conditional distributions necessary to perform the mass conversion. Instead, in the PS-theo approach, we assumed a log-normal analytical approximation of the conditional distributions of the concentrations with a median concentration-mass relation estimated from the Uchuu halo catalogues (see Eq.~2 in \citetalias[][]{Ishiyama_2021} with parameters set to the values given in Table 2 for the case `Fit all haloes')) with a width parameter $\sigma_{\log{c_{\rm vir}}} = 0.25$. These log-normal distributions were then converted into the NFW-sparsity conditional distributions using Eq.~(\ref{eqn:nfw_sparsity_concentration_stat}). Finally, in the case of the PD method, we simply computed the NFW-sparsities obtained assuming the Uchuu median concentration-mass relation from \citetalias{Ishiyama_2021}. 

In Fig.~\ref{fig:uchuu_pdfs}, we plot an example of these conditional distributions in the case of $s_{\rm vir,200}$ (solid lines) and $s_{\rm vir,500}$ (dashed lines), which for illustrative purposes we limit to three different bins of $M_{\rm vir}$ (panels from left to right) at $z\in\{0.0,0.14,0.25\}$ (panels from top to bottom), respectively. In each panel, the different distributions correspond to the NFW-inferred sparsities (PS-num PDF, brown curves), the `true' ones obtained from the ratio of the halo masses in the catalogues (NPS PDF, goldenrod curves), and the NFW-inferred sparsity value adopted in the deterministic approach (PD, pink vertical line) given by Uchuu's concentration-mass relation from \citetalias{Ishiyama_2021} at the central mass bin value. 

Firstly, by comparing the PS-num and NPS distributions for a given sparsity configuration, we can see that there is a systematic difference between the location of the peak of the NFW-inferred sparsity distribution and that of the true halo sparsity (i.e. obtained from the ratio of halo masses in the Uchuu catalogue). In particular, in the case of the NFW-sparsities, the distribution (brown curves) is shifted towards larger sparsity values than the true one (goldenrod curves) independently of mass, redshift, and overdensity considered. Secondly, we may notice that also the tails of the distribution are underestimated compared to those of the true sparsity distribution. Both these systematic differences appear to be of greater amplitude for $\Delta_2=500$ than $200$, and slightly increase with redshift. They result from the fact that the NFW profile provides a biased description of the logarithmic slope of the halo mass profile over different radial intervals, which correlates with the dynamical state of the haloes (Corasaniti et al, in prep.). Hence, we can expect such differences to lead to systematic effects on the converted HMFs and therefore on the predicted number counts. 

We stress the importance of the choice of the binning used to estimate the sparsity distribution when computing Eq.~(\ref{eq:nonparamstoc_conv}), since it has a significant impact on the accuracy of the results. This is because the sparsities of the halo population strongly vary with mass and redshift. As such, the binning must be adjusted accordingly. If the binning is too large, this may smooth out features of the underlying distribution, particularly at the high-mass end where the HMF varies rapidly. Conversely, if the binning is too narrow, it could introduce strong variations due to the inherent noise of the sparsely sampled distribution. To address this issue, we used a Bayesian adaptive binning algorithm based on Knuth's rule \citep{2006physics...5197K} implemented in the \texttt{astropy} library. This method accurately determines the optimal uniform bin-width to account for the mass and redshift dependence of the sparsity distributions.

\section{Cluster number counts}\label{ClusterCounts}
Let us consider a survey characterised by sky coverage $\Delta\Omega$ and selection $W(M_{\Delta},z)$; that is, the probability of detecting a cluster of mass $M_{\Delta}$ at redshift $z$. Then, the number of clusters in a given redshift bin of size $\Delta z$ centred at $z$ can be computed from the HMF as
\begin{equation}\label{eqn:numbercounts_gen}
    N(z) = \Delta\Omega \int_{z-\frac{\Delta z}{2}}^{ z+\frac{\Delta z}{2}}{\rm d}z\frac{{\rm d}^2V}{{\rm d}\Omega\, \diff z}\int {\rm d}M_\Delta\frac{{\rm d}n}{{\rm d}M_\Delta}\left(M_{\Delta},z\right)W(z,M_\Delta)\;,
\end{equation}
where
\begin{equation}\label{volume_element}
    \frac{{\rm d}^2V}{{\rm d}z\,{\rm d}\Omega} = \frac{c}{H(z)}(1+z)^2d_A^2(z)
\end{equation}
is the comoving volume element in solid angle ${\rm d}\Omega$ and redshift interval ${\rm d}z$, and $d_A$ is the angular diameter distance 
\begin{equation}
    d_A(z) = \frac{c}{H_0}\int_0^{z}\frac{{\rm d}z'}{E(z')}\;.
\end{equation}
In estimating the number counts, we assumed narrow redshift bins of size $\Delta{z}=0.1$, a sky coverage of $15\,000$ deg$^2$, and a mass-limited Heaviside selection function, $W(z,M_{\Delta})=\Theta(M_{\Delta}-M_{\Delta}^{\rm min})$. In such a case, Eq.~(\ref{eqn:numbercounts_gen}) reduces to a simpler form:
\begin{equation}\label{cluster_count}
    N(z) = \Delta\Omega\,\Delta z\frac{{\rm d}^2V}{{\rm d}\Omega \,{\rm d}z}\int_{M_{\Delta}^{\rm min}}^\infty {\rm d}M_\Delta\frac{{\rm d}n}{{\rm d}M_\Delta}\left(M_\Delta,z\right).
\end{equation}
\section{Cluster number counts data covariance}\label{data_cov}
We decomposed the covariance of galaxy cluster number count measurements as
\begin{equation}
    \mathbf{C}_\Delta = \mathbf{C}^{\rm S} + \mathbf{C}^{\rm SN}\;,
\end{equation}
where $\mathbf{C}^{\rm S}$ is the sample covariance and $\mathbf{C}^{\rm SN}$ is the shot noise covariance, which both depend on the specific mass definition ($\Delta = 200$ or $\Delta = 500$). We computed an unbiased estimate of the data covariance matrix using $10^3$ bootstrap iterations to generate mock realisations of the Flagship WIDE light-cone, from which we computed the covariance between the $i$-th and $j$-th redshift bin
\begin{equation}
    C_{\Delta,ij} = \frac{1}{N_B-1}\sum_{b=1}^{N_B}\left[N_i^{(b)} - \Bar{N_i} \right]\left[N_j^{(b)} - \Bar{N_j} \right]\;,
\end{equation}
where $N_B$ is the number of independent realisations, $N_i^{b}$ is the number count in the $i$-th redshift bin for the $b$-th bootstrap iteration, and $\Bar{N_i}$ is the mean number count in $i$-th bin over the full bootstrap samples; that is
\begin{equation}
    \Bar{N_i}= \frac{1}{N_B}\sum_{b=1}^{N_B}N_i^{b}\;.
\end{equation}
The shot noise contribution to the covariance, $\mathbf{C}^{\rm SN}$, is a diagonal matrix containing the Poisson variance in each redshift bin
\begin{equation}
    C^{\rm SN}_{ij} = \delta_{ij}\Bar{N_i}\;.
\end{equation}
The inverse of the covariance matrix $\mathbf{C}^{-1}_{\Delta}$ in  Eq.~(\ref{eq:likelihood_flagship}) is the precision matrix, which we corrected for the bias introduced by the finite number of bootstrap samples. In particular, we used the unbiased estimator of the precision matrix given by \citet{10.1093/mnras/stt270},
\begin{equation}
    \widehat{\mathbf{C}_\Delta^{-1}} =\left( \frac{N_B-N_z-2}{N_B-1}\right)\mathbf{C}_\Delta^{-1}\;,
\end{equation}
where $N_z$ is the number of redshift bins.

\section{Parameter degeneracies}\label{S8_degeneracy}

\subsection{Uchuu data analysis}
\begin{figure*}[ht]
    \centering
    \begin{minipage}{0.8\textwidth}
    \centering
        
        \begin{minipage}{0.45\textwidth}
            \centering
            \includegraphics[width=\linewidth]{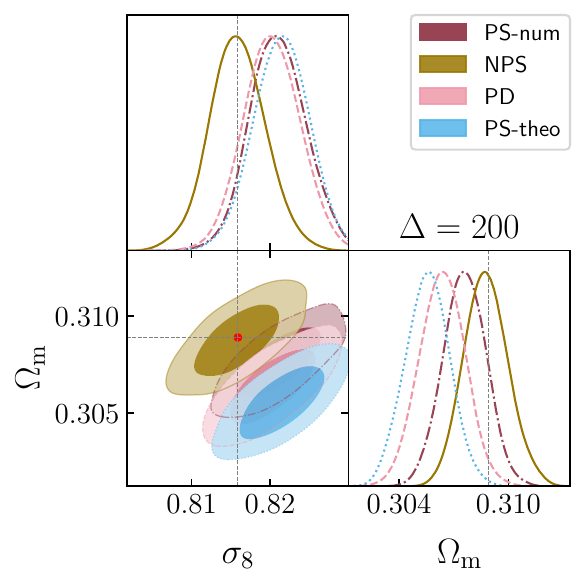}
        \end{minipage}
        \hfill
        \begin{minipage}{0.45\textwidth}
            \centering
            \includegraphics[width=\linewidth]{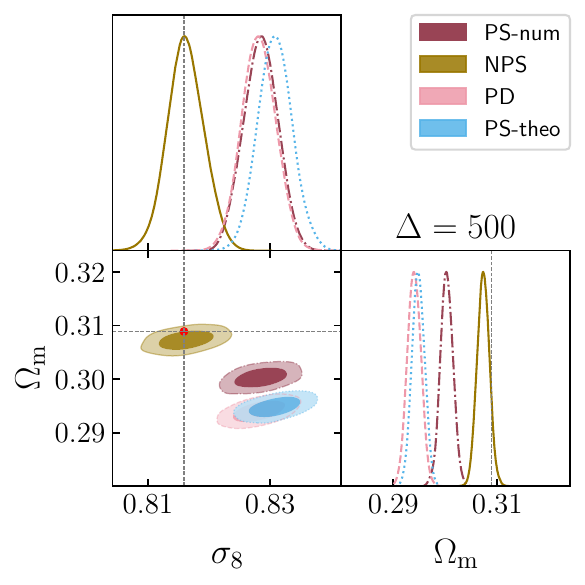}
        \end{minipage}
    \end{minipage}
    \caption{Triangular plot of the $1D$ and $2D$ marginalised posterior distributions on $\Omega_{\rm m}$ and $\sigma_8$ obtained from the analysis of Uchuu synthetic data at $M_{200}$ ({\it left panel}) and $M_{500}$ ({\it right panel}) with low-mass cut ($M_{\rm \Delta} \geq 3\times10^{13}\,M_{\odot}\,h^{-1}$ with $\Delta=200$ and $500$) assuming the {\it Euclid}\textbackslash Uchuu-HMF. The marginalised 1D posteriors correspond to the NPS (solid lines), PD (dashed lines), PS-num (dash-dotted lines), and PS-theo (dotted lines) mass conversion models. The shaded contours of the 2D marginalised posterior correspond to the $68\%$ and $95\%$ credible intervals for the NPS (goldenrod), PD (pink), PS-num (brown), and PS-theo (light blue) cases, respectively. The red dot in the 2D plot (vertical line in the 1D plot) corresponds to the Uchuu fiducial cosmological parameter value.}
    \label{fig:contour_3e13_Uchuu}
\end{figure*}

\begin{figure*}[ht]
    \centering
    \begin{minipage}{0.8\textwidth}
    \centering
            
        \begin{minipage}{0.43\textwidth}
            \centering
            \includegraphics[width=\linewidth]{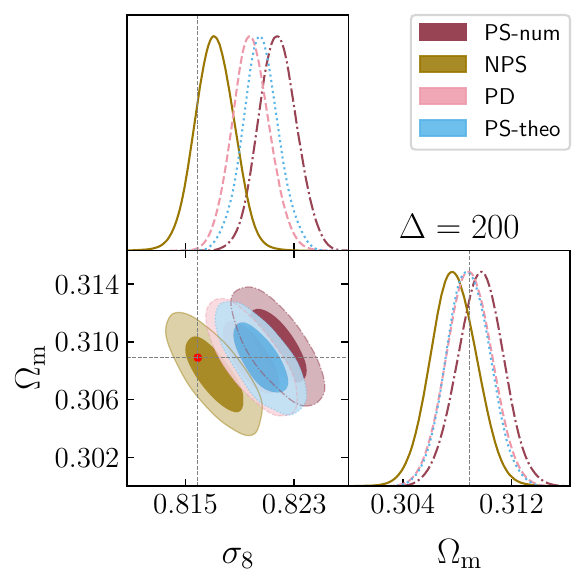}
        \end{minipage}
        \hfill
        \begin{minipage}{0.43\textwidth}
            \centering
            \includegraphics[width=\linewidth]{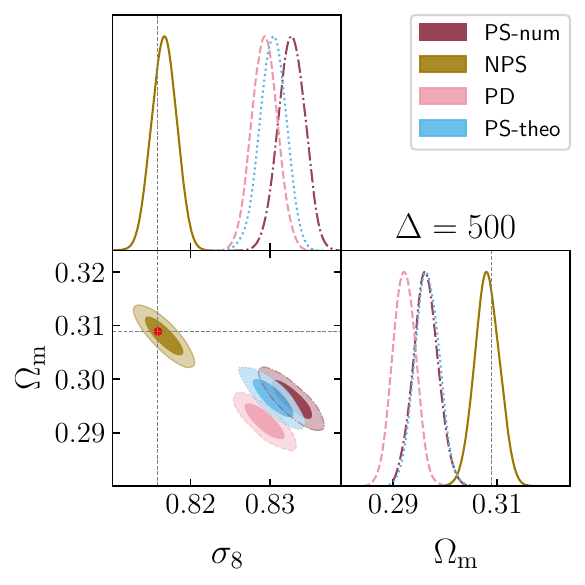}
        \end{minipage}
    \end{minipage}
    \caption{As in Fig.~\ref{fig:contour_3e13_Uchuu}, but the Uchuu synthetic data with high-mass cut ($M_{\rm \Delta} \geq 1\times10^{14}\,M_{\odot}\,h^{-1}$ with $\Delta=200$ and $500$).}
    \label{fig:contour_1e14_Uchuu}
\end{figure*}

\begin{figure*}[th]
\centering
\includegraphics[width = 0.43\linewidth]{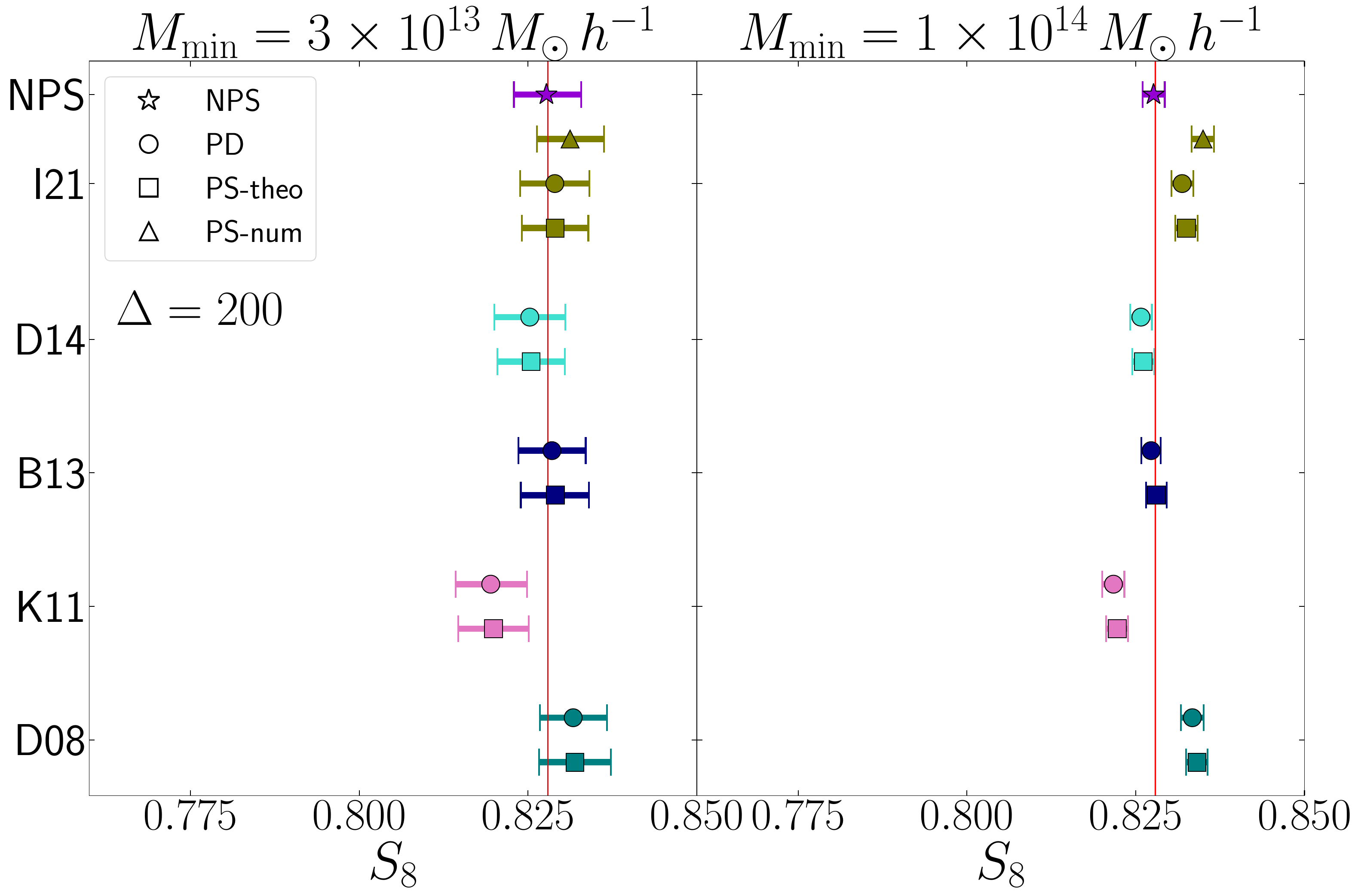}
\includegraphics[width = 0.43\linewidth]{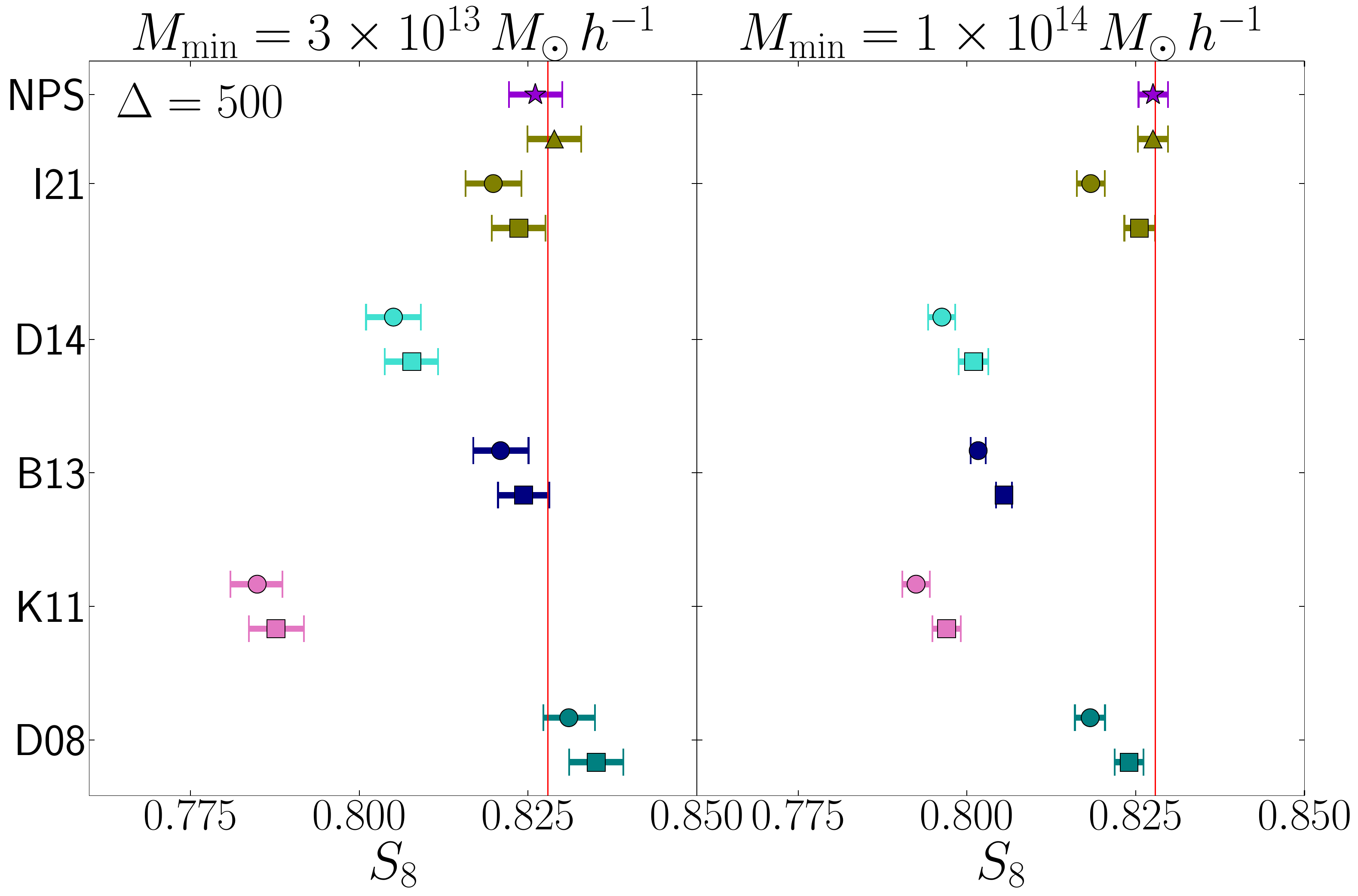}
\caption{Mean and standard deviation of the marginalised constraints on $S_8$ inferred from the analysis of Uchuu data at $M_{200}$ ({\it left panels}) and $M_{500}$ ({\it right panels}). In each panel, the left-hand (right-hand) side plot corresponds to the low (high) -mass cuts. The have been obtained by applying the mass conversion to the {\it Euclid}\textbackslash Uchuu-HMF calibrated at $M_{\rm vir}$ using the NPS (magenta star points), PD (filled circles), PS-num (filled triangles), and PS-theo (filled squares) mass conversion approaches. 
In the PD and PS-theo cases, we have assumed concentration-mass relation from the Uchuu dataset \citetalias{Ishiyama_2021} (olive green) and the relations from \citetalias{2014MNRAS.441.3359D} (cyan), \citetalias{2013ApJ...766...32B} (dark blue), \citetalias{2011ApJ...740..102K} (violet), and \citetalias{2008MNRAS.390L..64D} (dark green). The vertical lines show the fiducial value $S_8$.}
\label{fig:summary_plot_cM_S8}
\end{figure*}

Figures~\ref{fig:contour_3e13_Uchuu} and~\ref{fig:contour_1e14_Uchuu} show triangular plots of the 1 and 2D marginalised posteriors on $\Omega_{\rm m}$ and $\sigma_8$, inferred from the analysis of the Uchuu synthetic data samples with the low-mass cut and the high-mass cut, respectively, assuming the {\it Euclid}\textbackslash Uchuu-HMF mapped to $M_{200}$ (left panels) and $M_{500}$ (right panels) using the NPS, PS-num, PS-theo, and PD mass conversion approaches. As already mentioned in Appendix~\ref{ConditionalSparsityCalibration}, for the PS-theo and PD approaches, we assumed the Uchuu concentration-mass relation from \citetalias{Ishiyama_2021}, while in the PS-num case, we used the numerical distributions of halo concentrations from the Uchuu catalogues. The shaded areas of the 2D plots correspond to the $1$ and $2\sigma$ credible regions. As we can see, the confidence regions of the PD, PS-num and PS-theo cases are displaced with respect to the constraints from NPS along the degeneracy line of $\Omega_{\rm m}$ and $\sigma_8$. To account for the degeneracy, we inferred the constraint on the joint parameter $S_8=\sigma_8\sqrt{\Omega_{\rm m}/0.3}$. The mean and standard deviation for the different cases are shown in Fig.~\ref{fig:summary_plot_cM_S8}. Again, the NPS approach is the only method that always recovers the fiducial value. In the case of the PS-num, PS-theo, and PD methods, we still observe a systematic bias in the recovered value of $S_8$ depending on the target mass definition and the assumed concentration-mass relation. This is because the $S_8$ parameter does not accurately capture the parameter degeneracy of the cluster number count measurements.


\subsection{Flagship data analysis}\label{S8_degeneracy_flagship}

\begin{figure*}[ht]
    \centering
    \begin{minipage}{0.8\textwidth}
    \centering
        \begin{minipage}{0.45\textwidth}
            \centering
            \includegraphics[width=\linewidth]{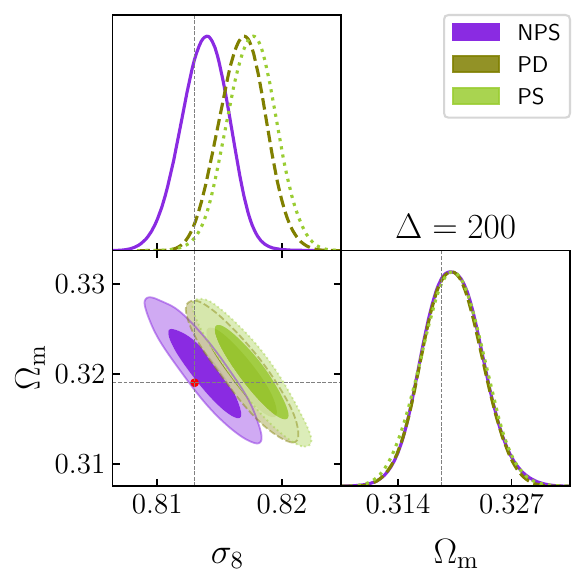}
        \end{minipage}
        \hfill
        \begin{minipage}{0.45\textwidth}
            \centering
            \includegraphics[width=\linewidth]{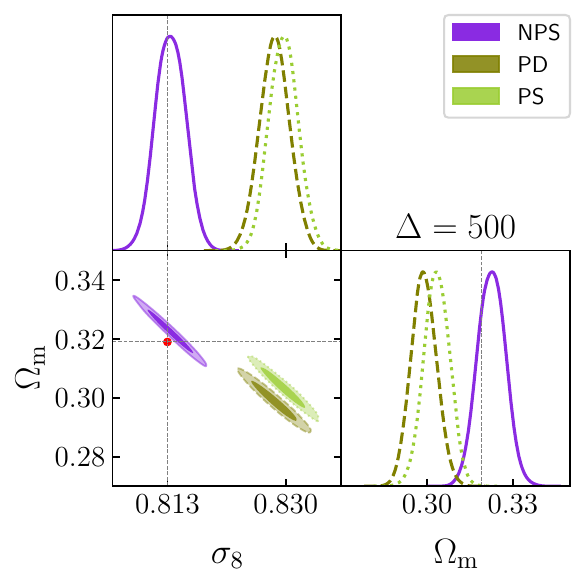}
        \end{minipage}
    \end{minipage}
    \caption{Triangular plot of the $1D$ and $2D$ marginalised posterior distributions on $\Omega_{\rm m}$ and $\sigma_8$ obtained from the analysis of Flagship synthetic data at $M_{200}$ ({\it left panel}) and $M_{500}$ ({\it right panel}) with $z_{\rm max}=1.2$ assuming the {\it Euclid}\textbackslash Uchuu-HMF mass converted using the NPS (magenta), PD (dark green), and PS (light green) methods assuming the Uchuu concentration-mass relation from \citetalias{Ishiyama_2021}. The red dot in the 2D plot (vertical line in the 1D plot) corresponds to the Flagship fiducial cosmological parameter value.}
    \label{fig:contour_Flagship}
\end{figure*}

\begin{figure*}[th]
\centering
\includegraphics[width = 0.45\linewidth]{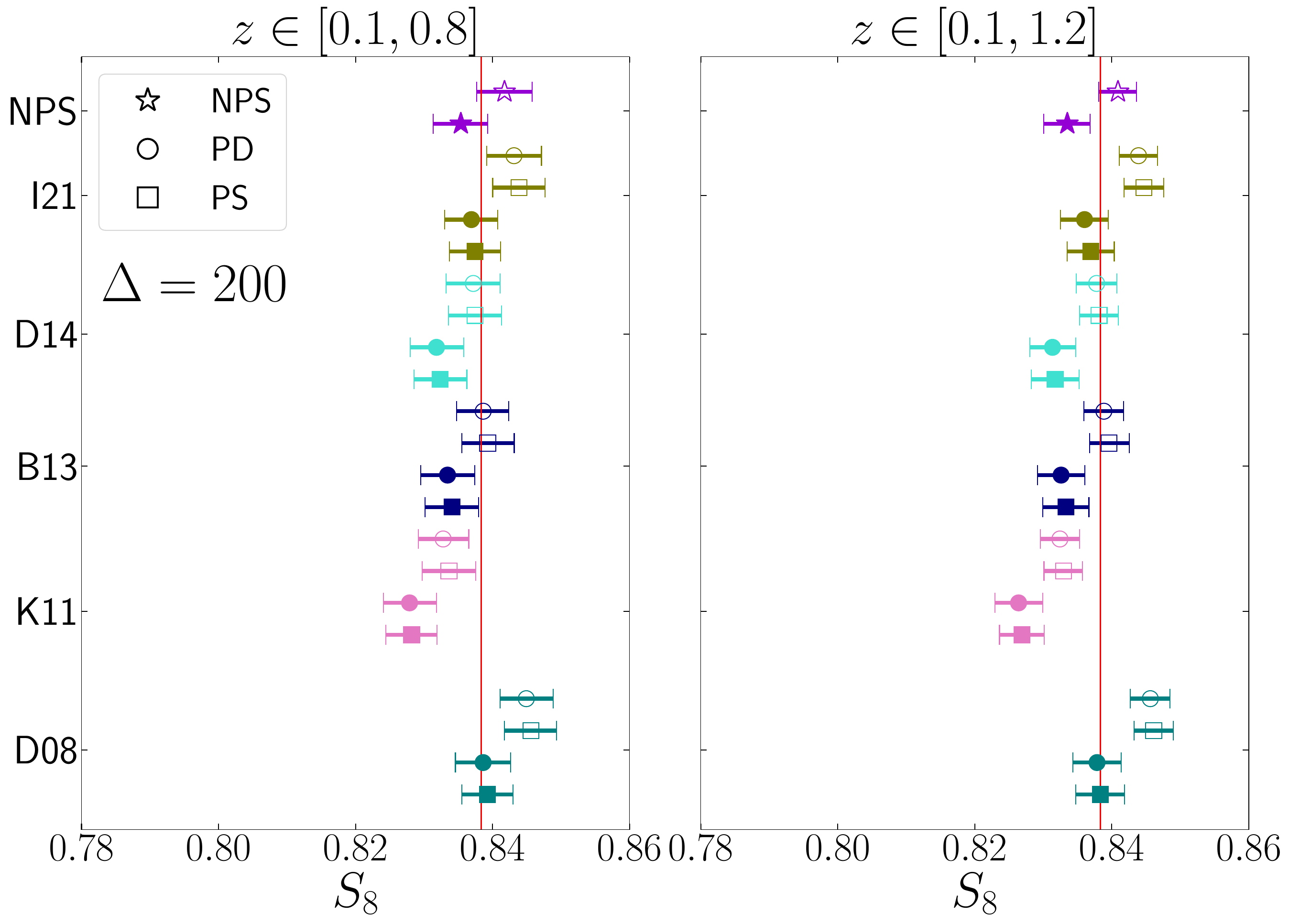}
\includegraphics[width = 0.45\linewidth]{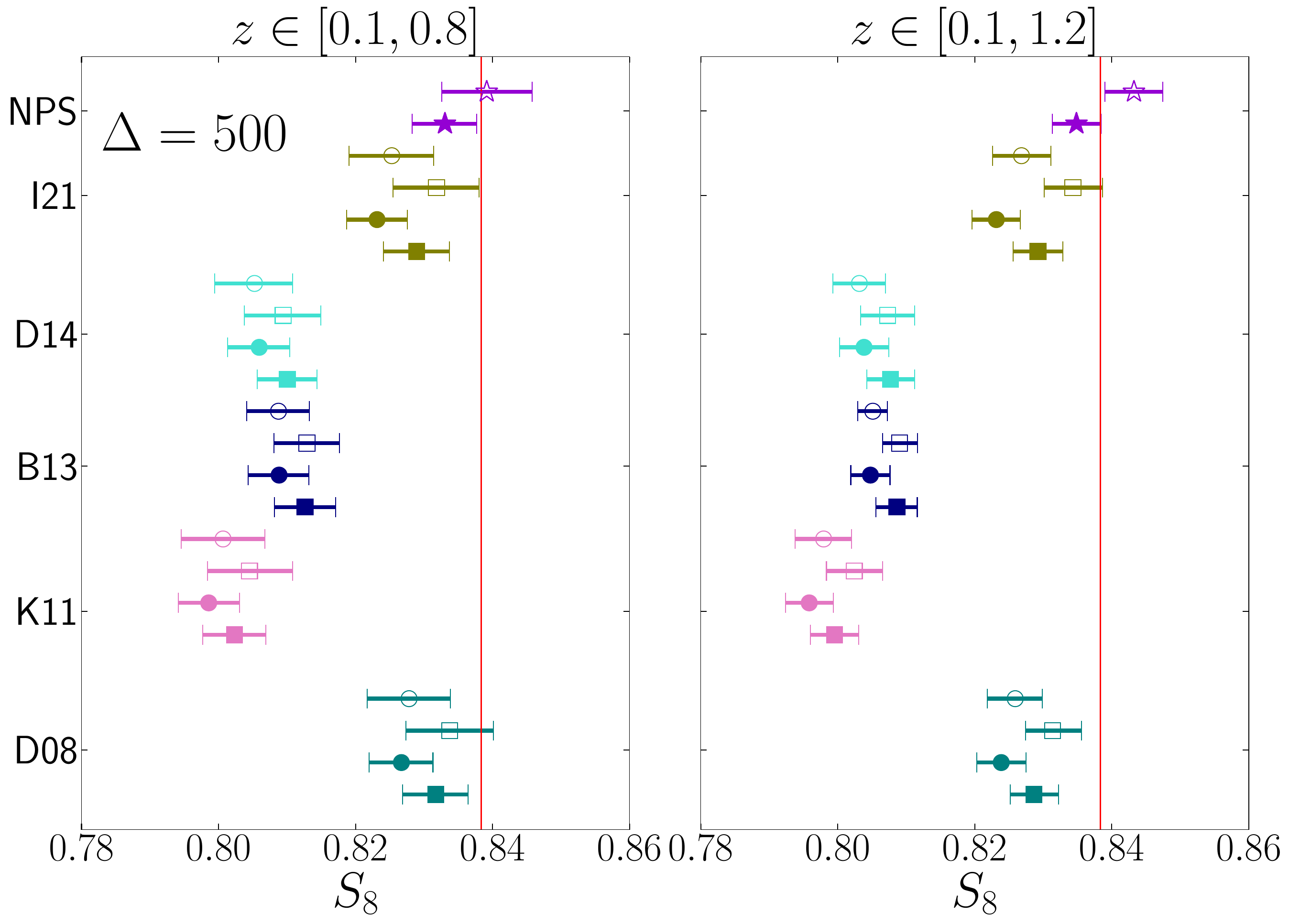}
\caption{Mean and standard deviation of the marginalised constraints on $S_8$ inferred from the analysis of Flagship data at $M_{200}$ ({\it left panels}) and $M_{500}$ ({\it right panels}). In each panel, the left-hand (right-hand) side corresponds to the low (high) redshift cut. These have been obtained using the NPS mass conversion (magenta stars), the PD (circles), and PS (squares) approaches for different concentration-mass relations as in Fig.~\ref{fig:summary_plot_cM_omegam_flagship}. The empty (filled) symbols correspond to the constraints inferred assuming the {\it Euclid}\textbackslash Uchuu-HMF ({\it Euclid}-HMF) mapped to the target mass definition of the Flagship datasets. The vertical line shows the fiducial value of $S_8$.}
\label{fig:summary_plot_cM_S8_zcuts}
\label{LastPage}
\end{figure*}

In Fig.~\ref{fig:contour_Flagship}, we show the triangular plots of the 1 and 2D marginalised posteriors on $\Omega_{\rm m}$ and $\sigma_8$, inferred from the analysis of the Flagship synthetic data with $z_{\rm max}=1.2$ assuming the {\it Euclid}\textbackslash Uchuu-HMF mass converted to $\Delta=200$ (left panel) and $\Delta=500$ (right panel) using the NPS approach, the PS, and PD methods assuming the Uchuu concentration-mass relation from \citetalias{Ishiyama_2021}. As in the case of the Uchuu data analysis, we can see the constraints from the PD and PS approaches are shifted along lines that are parallel to the major axis of the NPS ellipses. Nevertheless, when converted into constraints on the $S_8$ parameter as shown in Fig.~\ref{fig:summary_plot_cM_S8_zcuts}, we still find discrepancies with respect to the fiducial $S_8$ value that are on the same level of those obtained on $\Omega_{\rm m}$ and $\sigma_8$, and that depend on the target mass, the redshift interval of the synthetic data sample, and the assumed concentration-mass relation. This is mainly because $S_8$ does not accurately parametrise the parameter degeneracy of the cluster number counts.

\end{appendix}

\end{document}